\documentclass[12pt]{article}

\usepackage{amsmath}
\usepackage{amsfonts}
\usepackage{amssymb}
\usepackage{amsthm}
\input xy
 \xyoption{all}

\usepackage{fancyhdr}
\numberwithin{equation}{section}
\newcommand{\bs}{\boldsymbol}

\newtheorem{defi}{Definition}
\newtheorem{lemma}{Lemma}
\newtheorem{theo}{Theorem}

\newtheorem{example}{Example}

\newtheorem{corol}{Corollary}
\newcommand{\dt}{d_{\nabla}}
\newcommand{\dts}{d_{\nabla }^{\ast}}
\newcommand{\du}{d_{\nabla^{\mc{U}}}}

\renewcommand{\iota}{{\bf 1}}
\newcommand{\be}{\begin{equation}}
\newcommand{\ee}{\end{equation}}
\newcommand{\ba}{\begin{eqnarray}}
\newcommand{\ea}{\end{eqnarray}}

\newcommand{\mc}{\mathcal}

   \newcommand{\bul}{{\bullet}}

    \newcommand{\pfr}{\ar@<0.3ex>[r]}
  \newcommand{\pfl}{\ar@<0.3ex>[l]}
  \newcommand{\pfu}{\ar@<0.3ex>[u]}
  \newcommand{\pfd}{\ar@<0.3ex>[d]}

  \newcommand{\upflu}{\ar@<0.3ex>@{.>}[u]}
  \newcommand{\upfru}{\ar@<-0.3ex>@{.>}[u]}
  \newcommand{\upfld}{\ar@<-0.3ex>@{.>}[d]}
  \newcommand{\upfrd}{\ar@<0.3ex>@{.>}[d]}

  \newcommand{\bigupflu}{\ar@<0.3ex>[uu]}
  \newcommand{\bigupfru}{\ar@<-0.3ex>[uu]}
  \newcommand{\bigupfld}{\ar@<-0.3ex>[dd]}
  \newcommand{\bigupfrd}{\ar@<0.3ex>[dd]}

 \newcommand{\uoberer}{\bul \pfu \ar@<0.3ex>@{.>}[d] \pfl \pfr}

 \newcommand{\nuur}{\raisebox{2.5ex}{
                    \xymatrix@=2ex {
                     \bul \ar[r]      & \bul \\
                     \bul \ar@{.>}[u] & \bul } }}
  \newcommand{\nuru}{\raisebox{2.5ex}{
                    \xymatrix@=2ex {
                     \bul        & \bul \\
                     \bul \ar[r] & \bul \ar@{.>}[u]} }}
  \newcommand{\nuld}{\raisebox{2.5ex}{
                    \xymatrix@=2ex {
                     \bul \ar@{.>}[d] & \bul \ar[l] \\
                     \bul             & \bul} }}
  \newcommand{\nudl}{\raisebox{2.5ex}{
                    \xymatrix@=2ex { 
                     \bul & \bul \ar@{.>}[d] \\
                     \bul & \bul \ar[l]} }}
  \newcommand{\nudlur}{\raisebox{2.5ex}{
                      \xymatrix@=2ex {
                       \bul \ar@{.>}[d] & \bul \ar[l] \\
                       \bul \ar[r]      & \bul \ar@{.>}[u]} }}
  \newcommand{\nuldru}{\raisebox{2.5ex}{
                      \xymatrix@=2ex {
                       \bul \ar[r]      & \bul \ar@{.>}[d] \\
                       \bul \ar@{.>}[u] & \bul \ar[l]} }}
  \newcommand{\nud}{\raisebox{2.5ex}{
                     \xymatrix@=2ex {
                      \bul \ar@{.>}[d] & \bul   \\
                      \bul             & \bul} }}
  \newcommand{\nuu}{\raisebox{2.5ex}{
                     \xymatrix@=2ex {
                      \bul             & \bul   \\
                      \bul \ar@{.>}[u] & \bul} }}
  \newcommand{\nuud}{\raisebox{2.5ex}{
                     \xymatrix@=2ex {
                      \bul \ar@<-0.3ex>@{.>}[d] & \bul   \\
                      \bul \ar@<-0.3ex>@{.>}[u] & \bul} }}
  \newcommand{\rnudl}{\raisebox{2.5ex}{
                    \xymatrix@=2ex {
                        \bul \ar[r] & \bul \ar@{.>}[d] \\
                        \bul        & \bul \ar[l]}    }}
  \newcommand{\rnuul}{\raisebox{2.5ex}{
                      \xymatrix@=2ex {
                        \bul        & \bul \ar[l]      \\
                        \bul \ar[r] & \bul \ar@{.>}[u]} }}
  \newcommand{\rnuunudl}{\raisebox{2.5ex}{
                         \xymatrix@=2ex {
                          \bul                & \bul \ar@<-0.3ex>@{.>}[d] \\
                          \bul \ar@<-0.3ex>[r] & \bul \ar@<-0.3ex>[l] \ar@<-0.3ex>@{.>}[u]} }}
  \newcommand{\dlur}{\raisebox{2.5ex}{
                      \xymatrix@=2ex {
                       \bul \ar[d] & \bul \ar[l] \\
                       \bul \ar[r] & \bul \ar[u]} }}
  \newcommand{\dlurklein}{\raisebox{2.5ex}{
                           \xymatrix@=1.5ex {
                            \bul \ar[d] & \bul \ar[l] \\
                            \bul \ar[r] & \bul \ar[u]} }}
 
  \newcommand{\ldru}{\raisebox{2.5ex}{
                      \xymatrix@=2ex {
                       \bul \ar[r] & \bul \ar[d] \\
                       \bul \ar[u] & \bul \ar[l]} }}
  \newcommand{\ldur}{\raisebox{2.5ex}{
                      \xymatrix@=2ex {
                       \bul \ar@<0.3ex>[d] \ar@<0.3ex>[r] & 
                       \bul \ar@<0.3ex>[l]  \\
                       \bul \ar@<0.3ex>[u] } }}
  \newcommand{\dlru}{\raisebox{2.5ex}{
                      \xymatrix@=2ex {
                       & \bul \ar@<0.3ex>[d] \\
                       \bul \ar@<0.3ex>[r] & 
                       \bul \ar@<0.3ex>[l] \ar@<0.3ex>[u]} }}
        \newcommand{\tur}{\raisebox{2.5ex}{
                    \xymatrix@=1ex {
                      \ar[rr]   & &  \\
                                & &  \\
                      \ar[uu]   & &     }}}

  \newcommand{\tru}{\raisebox{2.5ex}{
                    \xymatrix@=1ex {
                                & &  \\
                                & &  \\
                      \ar[rr]   & &  \ar[uu]   }}}

\newcommand{\tld}{\raisebox{2.5ex}{
                    \xymatrix@=1ex {
                      \ar[dd]   & &  \ar[ll]  \\
                                & &           \\
                                & &     }}}

  \newcommand{\tdl}{\raisebox{2.5ex}{
                    \xymatrix@=1ex {
                                & &  \ar[dd]  \\
                                & &  \\
                                & &  \ar[ll]  }}}

  \newcommand{\tnoneunitaryur}{\raisebox{2.5ex}{
                              \xymatrix@=1ex {
                              \ar[rr]        & &  \\
                                             & &  \\
                              \ar@{.>}[uu]   & &     }}}

 \newcommand{\tnoneunitaryru}{\raisebox{2.5ex}{
                              \xymatrix@=1ex {
                                             & &  \\
                                             & &  \\
                              \ar[rr]        & &  \ar@{.>}[uu]   }}}

 \newcommand{\tnoneunitaryld}{\raisebox{2.5ex}{
                              \xymatrix@=1ex {
                              \ar@{.>}[dd]   & &  \ar[ll] \\
                                             & &           \\
                                             & &     }}}

 \newcommand{\tnoneunitarydl}{\raisebox{2.5ex}{
                              \xymatrix@=1ex {
                                  & &  \ar@{.>}[dd]  \\
                                  & &                \\
                                  & &  \ar[ll]   }}}

 \newcommand{\tnoneunitaryud}{\raisebox{2.5ex}{
                              \xymatrix@=1ex {
                               \ar@<0.3ex>@{.>}[dd]   &   \\
                                                      &   \\
                               \ar@<0.3ex>@{.>}[uu]   &   }}}

\newcommand{\tnoneunitarydu}{\raisebox{2.5ex}{
                              \xymatrix@=1ex {
                               \ar@<-0.3ex>@{.>}[dd]   &   \\
                                                       &   \\
                               \ar@<-0.3ex>@{.>}[uu]   &   }}}

 \newcommand{\turdl}{\raisebox{2.5ex}{
                     \xymatrix@=1ex{
                           \ar[rr] & &  \ar@{.>}[dd]  \\
                                   & &           \\
                           \ar@{.>}[uu] & &  \ar[ll]  }}}

   \newcommand{\tturdl}{\raisebox{2.5ex}{
                     \xymatrix@=1ex{
                           \ar[rr] & &  \ar[dd]  \\
                                   & &           \\
                           \ar[uu] & &  \ar[ll]  }}}

\def\rellow#1#2{Mathrel{Mathop{\kern 0pt #1}\limits_{#2}}}

\title{Generalized Gauge Theories with Nonunitary Parallel Transport:\\
General Relativity with Cosmological Constant as an Example
  }
\author{Gerhard Mack and Thorsten Pr\"ustel\\
   II. Institut f\"ur Theoretische Physik, Universit\"at Hamburg \footnote{e-mail: gerhard.mack@desy.de,
                                                                                                                                          thorsten.pruestel@desy.de}
\date{\today}}
\begin{document}

\maketitle
$\quad$\\[-5mm]\noindent 

\begin{abstract}
  In gauge theories parallel transporters (PTs) $\mc{U}(C)$ along paths
  $C$ play an important role. Traditionally they are unitary or
  pseudoorthogonal maps between vector spaces.  We propose to abandon
  unitarity of parallel transporters and with it the {\em a priori} 
  assumption of  metricity in general relativity.
  A $\ast$-operation on parallel transporters serves as a
  substitute for it, and this $\ast$-operation is proven to be unique
  on group theoretical grounds.  The vierbein and the spin connection
  appear as distinguishable parts of a single de Sitter gauge field
  with field strength $\bs{F}$. The action takes the form
  $\frac{3}{16\pi
    G\Lambda}\int\text{tr}(\bs{F}\wedge\bs{F}i\bs{\gamma}_{5})$ and
  both the Einstein field equations with arbitrarily small but
  nonvanishing cosmological constant $\Lambda$ and the condition of vanishing
  torsion are obtained from it.  The equation of motion for classical
  massive bodies turns out to be de Sitter covariant.  %Finally we
  %consider an extension of general relativity by considering a
  %conformal holonomy group.
\end{abstract}
\newpage %%%%%%%%%%%%%%%%%%%%%%%%%%%%%%%%%%%%%%%%%%%%%%%
\section{Introduction}\label{sec:1} 
General relativity and the gauge theories governing the dynamics of elementary particles obey very much the same basic principles, yet they are different both in their variables and in their action. More precisely, apart from the vector potential (=spin connection) a vierbein field appears in general relativity
 which has no analogue in Yang-Mills theories, and the Einstein-Hilbert action is linear in the curvature, while the Yang-Mills action is quadratic in the field strength. 

The Kaluza Klein principle addresses this issue by constructing gauge fields from metric tensors through dimensional reduction. In this approach, the properties of pseudo-Riemannian
space time manifolds are assumed basic.
We wish to avoid such assumptions a priori beyond the fundamental locality properties of gauge theories.

General relativity and gauge theories demonstrate that strong restrictions flow from the requirement that fundamental equations must be meaningful given the assumed a priori
structure. In general relativity, the assumed a priori structure includes neither preferred coordinate systems nor the possibility of comparing directions in different fibres $
T_{x}\mc{M}$ of tangent space. This suggests the strategy of lessening what is assumed as a priori structure, rather than adding to it \cite{Mack:1997hx}. In this paper we propose to abandon the a priori assumption of metricity
in general relativity, substituting for it the existence of a $\ast$-operation on parallel transporters. We prove that a nontrivial $\ast$-operation is unique, if the holonomy group is a de Sitter group.
The lesson is that one can start from a single de Sitter gauge field $\bs{B}_{\mu}(x)$. The $\ast$-operation determines a split of $\bs{B}_{\mu}(x)$ into a spin connection $\bs{A}_{\mu}(x)$ and a vierbein.
$\bs{A}_{\mu}(x)$ furnishes a metric connection with metric given by the vierbein.
 
 In all gauge theories, including general relativity, parallel transporters (PTs)
 \begin{equation}
 \mc{U}(C): V_{x}\rightarrow V_{y}
 \end{equation}
 along paths $C$ between points $x,y$ of the space time manifold $\mc{M}$ play a basic role. Traditionally, one demands that parallel transport forth and back yields the identity, and that the fibres $V_{x}$ come equipped with a bilinear or sesquilinear form $\langle\quad,\quad\rangle_{x}$ which is preserved by parallel transport. In general relativity this is the assumption
 of metricity. Defining the adjoint $\mc{U}(C)^{\ast}$ by $\langle \mc{U}(C)^{\ast}v,w\rangle_{x}=\langle v, \mc{U}(C)w\rangle_{y}$, the stated demands read
 \begin{equation}\label{unitary}
 \mc{U}(-C)=\mc{U}(C)^{-1}=\mc{U}(C)^{\ast}
 \end{equation}
 where $-C$ is path $C$ traversed in the opposite direction. We refer to the second equality as unitarity of parallel transporters.
 
We propose to abandon the requirement of their unitarity and to retain only the existence of a $\ast$-operation such that
\begin{equation}
\mc{U}(C)^{\ast}: V_{y}\rightarrow V_{x}
\end{equation} 
is defined as a linear map with the properties
\begin{equation}
\begin{split}
(\mc{U}(C_{2})\mc{U}(C_{1}))^{\ast} & =\mc{U}(C_{1})^{\ast}\mc{U}(C_{2})^{\ast}\\
\mc{U}(\emptyset)^{\ast} & =\mc{U}(\emptyset)=\text{id}.\\
\end{split}
\end{equation}
 
 It was shown in earlier papers
 \cite{ThorstenDiss,Lehmann:2003jh,Lehmann:2003jn}
that this generalization leads to a geometric interpretation of Higgs fields. They can appear as parts of generalized parallel transporters 
 in extra directions in models in 5 (or more) dimensions in a novel way such that exponential mass hierarchies appear when the local gauge symmetry is spontaneously broken by a Higgs mechanism. 
 
 Assuming invertibility of PTs, a holonomy group $H$ can be defined which inherits the $\ast$-operation
 which in turn defines an involutive automorphism 
$\theta(g)=g^{\ast -1}$ of $H$. Conversely the involutive automorphism $\theta$ 
suffices to fix the $\ast$-operation on PTs modulo gauge transformations in $H$. Involutive automorphisms $\theta$ and $\theta'$ which are conjugate in $H$ (viz. $\theta'(g)=g_{1}\theta(g)g_{1}^{-1}$ for some $g_{1}\in H$
and all $g\in H$) are not essentially different. They are related by gauge transformations in $H$.

 Assuming the holonomy group is a Lie group (or dense in a Lie group)
 $H$, its conjugacy classes of involutive automorphisms can be classified. There are few possibilities. Moreover, there is a distinguished subgroup $G$ consisting of elements $u\in H$ obeying $u^\ast = u^{-1}$. 
  
For general relativity, the holonomy group $H$ is postulated to be a  de Sitter group $SO(1,4)$ or $SO(2,3)$ (or rather their simply connected covering groups). The second possibility $SO(2,3)$ is distinguished by admitting chiral fermions.
But $SO(1,4)$ is favored by the experimental fact that the cosmological constant is positive. In both cases, a nontrivial $\ast$-operation leads to $G=SO(1,3)$, assuming $G$ is noncompact\footnote{A compact $G$ would lead to Riemannian rather than pseudo-Riemannian geometry and is incompatible with the assumption of a causal structure.}. This is the traditional gauge group of general relativity according to Utiyama \cite{Utiyama:1956sy}.
The $\ast$-operation is proven to be unique in this case, modulo gauge transformations. Generally speaking, there is de Sitter covariance, but general gauge transformations in $H$ transform the $\ast$-operation on parallel transporters. Once the $\ast$-operation is fixed, only $G$ survives as a local symmetry.

In this paper we abandon the assumption of metricity, but parallel transporters remain invertible. This is implied by the assumed existence of vector potentials which we retain for now. The consideration of noninvertible parallel transporters is outside the scope of this paper, although they could be physically interesting for gauge theories in space time with defects.

The $\ast$-operation identifies vierbein and spin connection as parts of a single de Sitter vector potential. We propose an action for this vector potential which has a $(curvature)^{2}$-form. Its variation leads to the Einstein field equations with an arbitrarily small but nonvanishing cosmological constant. It is intrinsically positive if $H=SO(1,4)$.
 
 The paper is organized as follows.
 In section \ref{polarsection}  we present the framework of generalized gauge theories,  proving the above mentioned classification and uniqueness results for  $\ast$-operations. We discuss in section \ref{diracspinors} the generalized parallel transport of Dirac and Weyl spinors, respectively. Vierbein and spin connection become identifyable pieces of a single de Sitter vector potential. In section \ref{genmetr} we show that there is a canonical way of constructing a metric and a metric connection
 (with unitary parallel transporters $\mc{U}(C)$).
 % Covariant derivatives of endomorphism valued forms will be introduced in section \ref{covariantderivatives}.% 
In section \ref{gravityaction} we propose an action which takes the  $(curvature)^{2}$ form and derive equations of motion for it. One may adjoin the involutive automorphism as a group element $\Theta$ to $H$, it is represented by $\bs{\Theta}=-i\bs{\gamma}_{5}$
in the Dirac spinor representation. The proposed action reads $S=\frac{1}{g^{2}}\int\text{tr}(\bs{F}\wedge\bs{F}^{\ast}\bs{\Theta})=-\frac{1}{g^{2}}\int\text{tr}(\bs{F}\wedge\bs{F}\bs{\Theta})$ where $\bs{F}$ is the de Sitter field strength in Dirac spinor representation. Variation leads to Einstein-Palatini field equations with gravitational constant $G$
and cosmological constant $\Lambda\neq 0$ when one identifies $\frac{1}{g^{2}}=\frac{3}{16\pi G\Lambda}$. 

The field equations can be written in a compact form as follows. 
Denote the de Sitter parallel transporter by $\mc{T}$. 
Thanks to the possibility of substituting $\mc{T}(C)^{\ast -1}
= \Theta(\mc{T}(C))$ for 
$\mc{T}(C)$, there are
actually two different\footnote{There are also two field strengths, but they
 are related as $\bs{F}$ and $-\bs{F}^\ast$} exterior covariant derivatives,
denoted $\dt$ and $\dts$, acting on 
endomorphism valued forms such as $\bs{F}$. They are conjugate under 
$\bs{\Theta}$ in the sense that 
\be \dt (\bs{F} \bs{\Theta}) = (\dts\bs{F})\bs{\Theta} \label{thetaConjug}
\ee
 We have $\dt \bs{F}=0$
as a Bianchi identity, whereas variation of $S$ leads to the field equations
\be \dts (\bs{F-F^\ast})=0\ , \ee
or, equivalently, $\dt \left( \bs{F}\bs{\Theta} -
 \bs{F}^\ast \bs{\Theta}\right)=0$. 

It will be shown in section \ref{gravityaction} that our action differs 
from the Einstein-Palatini action by a topological term.
One may  add a  further topological term 
$\propto \int\text{tr}(\bs{F}\wedge\bs{F})$ to
 the action. In any case, the action is polynomial. This is not a surprise. 
Similar actions have been considered in the literature before by MacDowell and
 Mansouri \cite{MacDowell:1977jt}, and Smolin \cite{Smolin:1998qp}. Smolin and Starodubtsev \cite{Smolin:2003qu} 
 also considered what happens when one substitutes a dynamical fields for 
$\bs{\Theta}$. Freidel and Starodubtsev \cite{Freidel:2005ak} argued that 
general
 relativity with a cosmological constant becomes renormalizable when one 
treats it as a perturbation 
of a topological field theory whose partition function is to be 
evaluated exactly.

 Finally we  show  in section \ref{classical} that  classical massive bodies can be treated within the de Sitter theory as well.

Attempts to regard vierbeins as gauge fields have been made before \cite{MacDowell:1977jt}. However, it was said that `` there has  always been  something contrived about attempst to
 interpret  general relativity as a gauge theory in that narrow sense''\cite{Witten:1988hc}. We hope that the present approach is more convincing. 

% Finally we study  extensions of general relativity in section \ref{conformalholonomy}. %%%%%%%%%%%%%%%%%%%%%%%%%%%%%%%%%%%%%%%%%%%%%%%%%%%%%%%%%%%%%%%%%%%%%%
 %%%%%%%%%%%%%%%%%%%%%%%%%%%%%%%%%%%%%%%%%%%%%%%%%%%%%%%%%%%%%%%%%%%%%%
\section{ Generalized parallel transport}\label{polarsection}
 %%%%%%%%%%%%%%%%%%%%%%%%%%%%%%%%%%%%%%%%%%%%%%%%%%%%%%%%%%%%%%%%%%%%%%
 \subsection{Holonomy group}
\label{holonomy}
Let us consider a gauge theory with possibly nonunitary parallel transporters $\mc{T}(C)$ along paths $C$.

There are two stages to the generalization 
\begin{itemize}
\item[i.)] The parallel transporters and their adjoints are invertible, but not necessarily unitary.
\item[ii.)] The inverses $\mc{T}(C)^{-1}$ may not exist at all.
\end{itemize}
Assuming invertibility, the fibres $V_x$ must have constant dimension.
\begin{defi}[Generalized connection]
Given a vector bundle over a differentiable manifold $\mc{M}$ with fibres $V_{x}$ at $x\in\mc{M}$, an invertible generalized connection $\mc{T}$ consists of an assignment of invertible maps $\mc{T}(C):V_{x}\rightarrow V_{y}$
to every piecewise smooth path $C$ from $x$ to $y$ such that the composition rules
\begin{equation}
\mc{T}(C_{2}\circ C_{1})=\mc{T}(C_{2})\mc{T}(C_{1}),\quad\mc{T}(\emptyset)=\text{id}
\end{equation}
hold, together with a $\ast$-operation which takes $\mc{T}(C)$ to an invertible map $\mc{T}(C)^{\ast}:V_{y}\rightarrow V_{x}$, and $\mc{T}(C)^{\ast}$ to $\mc{T}(C)$, such that
\begin{equation}
(\mc{T}(C_{2})\mc{T}(C_{1}))^{\ast}=\mc{T}(C_{1})^{\ast}\mc{T}(C_{2})^{\ast},\quad \mc{T}(\emptyset)^{\ast}=\mc{T}(\emptyset)=\text{id}.
\end{equation}
The generalized connection is differentiable if there exists for every $\hat x$ a moving frame $e_{\alpha}(x)$ in  a neighborhood $\mc{N}$ of $\hat x$ which furnishes a basis in $V_{x}$ for every $x\in\mc{N}$
 and such that the following is true.
 For paths $C:x\rightarrow y$ in $\mc{N}$parameterized by $\tau\in[0,\tau_{e}]$, define the parallel transport matrices $\bs{T}(C)$
 by
 \begin{equation}
 \mc{T}(C)e_{a}(x)=e_{b}(y)\bs{T}(C)^{b}\,_{a}.
 \end{equation}
Let $C_{t}$ be the piece of the path $C$ from $C(0)$ to $C(t),t<\tau_{e}$. Then $\bs{T}(C_{t})^{b}\,_{a}$ is differentiable with respect to $t$, for all smooth paths $C$ and all $\tau$.

Given a differentiable connection and the moving frame, a vector potential $\bs{B}_{\mu}(x)$ is defined,
\begin{equation}
Y^{\mu}\bs{B}_{\mu}(z)=-\frac{d}{dt}\bs{T}(C_{t}),
\end{equation}
where $z=C(t)$ and $Y$ is the tangent vector to $C$ at $z$.
\end{defi}
 A holonomy group may be defined as follows. 
\begin{defi}[Holonomy group]
Refer to parallel transporters $\mc{T}(C_i)$, their adjoints $\mc{T}(C_i)^\ast$ and inverses $\mc{T}(C_i)^{-1}, \mc{T}(C_i)^{\ast -1}$ as generalized parallel transporters (gPTs). They may be composed to PTs around closed loops 
$C=C_n\circ...\circ C_1$. The totality of all such PTs along loops $C$ from 
$x$ to $x$ form a group $H_x$. The isomorphism class $H$ of $H_{x}$ will be called 
the holonomy group for now.
\end{defi}
If $C^\prime$ is a path from $x$ to $y$ and 
$g\in H_x$ then $\mc{T}(C^\prime)g\mc{T}(C^\prime)^{-1} \in H_y$, and this defines a homomorphism from $H_x$ to $H_y$. The inverse homomorphism also exists. Therefore $H_x$ and $H_y$ are isomorphic to a group $H$ and $H$ does not depend on the choice of $x$. 
 Later we admit closure in a suitable topology so that $H$ becomes a Lie group.
  
A $\ast$-operation on PTs defines an involutive automorphism $\theta$ of the holonomy group,
\be \theta (g)= g^{\ast -1} \mbox{ for } g\in H_x  \label{invol} \ee
Involutive means $\theta^2=\text{id}$. 

With a holonomy group at hand, the structure of the theory remains similar to conventional gauge theories which can be handled with the method of principal fibre bundles. 
In particular, we may restrict attention to moving frames which are obtained by parallel transport of a basis at $\hat x$  along some paths. Then the parallel transport matrices take their values in a group of matrices which is a matrix representation of $H$. Therefore we may regard $H$ as a group of matrices and $\bs{T}(C)\in H$. This will be assumed from now on.

When invertibility is given up, the situation becomes much more complicated. 
For instance, space time $\mc{M}$ may decompose into domains $\mc{M}_{i}$ separated by boundaries, such that the parallel transporters along paths within domains remain invertible, defining holonomy groups $H_i$, but the PT across boundaries are not invertible so that gPTs along arbitrary loops $C:x\mapsto x$ define semigroups $S_i\supset H_i$ for $x\in \mc{M}_{i}$. Instead of a single holonomy group we now have collections of holonomy semigroups $S_i\supset H_i$ and PTs across boundaries form intertwiners between these. A comprehensive study of such possibilities is beyond the scope of this paper, although such theories may be physically interesting to describe physics on space time with defects.

In this paper we restrict attention to the situation where all  parallel transporters remain  invertible.
%%%%%%%%%%%%%%%%%%%%%%%%%%%%%%%%%%%%%%%%%%%%%%%%%%%%%%%%%%%%%%%%%%%%%%%%%%%%%%%%%%%%%%%%%
\subsection{Involutive automorphisms and polar decomposition of parallel transporters}
By definition, an active gauge transformation in $H$ is defined by a choice of $g_{z}\in H_{z}$ for all $z$, and PTs $\mc{T}(C)$ along paths $C$ from $x$ to $y$
transform into $g_{y}^{-1}\mc{T}(C)g_{x}$.
Specializing to $y=x$, the group $H_{x}$ gets transformed into itself.
Suppose the holonomy group is a Lie group or dense in a Lie group $H$ in a suitable topology such that group multiplication and adjunction are continuous. Henceforth $H$ will be called the holonomy group. Then $H$ inherits the involutive automorphism $\theta$ defined by eq.(\ref{invol}) and $\theta$ passes to an involutive automorphism, also denoted by $\theta$, of the (real) Lie algebra $\mathfrak{h}$ of $H$.  

We will see below that any  involutive automorphism of $H$ suffices to 
define the $\ast$-operation on PTs modulo active gauge transformations in $H$. Involutive automorphisms $\theta$ of $H$ are regarded as inequivalent if they are not conjugate within $H$. There are few possibilities of inequivalent involutive automorphisms of $H$  and they can be classified by pairs of real forms of the complexification of $\mathfrak{h}$ as follows.

\begin{theo}[Involutive automorphisms]\label{automorphisms}
Let $H$ be a Lie group with involutive automorphism $\theta$ which passes to  the Lie algebra $\mathfrak{h}$ of $H$. Let $G$ be the Lie subgroup of $H$ whose elements obey $\theta(g)=g$.
Regard $\mathfrak{h}$ as a real form of its complexification
  $\mathfrak{h}^\mathbb{C}$.

1. There exists another real form $\mathfrak{h}_{c}$ of   $\mathfrak{h}^\mathbb{C}$
such that the Lie algebra  of $G$ equals $\mathfrak{g}=\mathfrak{h}\cap \mathfrak{h}_{c}$.

2. There exists a decomposition of $\mathfrak{h}$ as a direct sum
\be \mathfrak{h}=\mathfrak{g}\oplus\mathfrak{p} \label{Cartan}
\ee
such that 
\be [\mathfrak{g},\mathfrak{g}]\subset \mathfrak{g}, \quad 
 [\mathfrak{p},\mathfrak{g}]\subset \mathfrak{p}, \quad
 [\mathfrak{p},\mathfrak{p}]\subset \mathfrak{g}\ . \label{CartanProp}
\ee
and the involutive automorphism $\theta$ acts according to 
\be \theta(X)=X \mbox{ for } X\in \mathfrak{g}, \quad
\theta(X)=-X \mbox{ for } X\in \mathfrak{p}, \label{thetaact}\ee
and 
\be \mathfrak{h}_{c} = \mathfrak{g}\oplus i\mathfrak{p}, \label{hprime} \ee
where $i=\sqrt{-1}$. 
\end{theo}
\begin{proof}[
{\sc Proof of Theorem \ref{automorphisms}.}] Given $\theta$, the decomposition (\ref{Cartan}) is defined by 
eqs.(\ref{thetaact}) and fulfills (\ref{CartanProp}). 
Defining $\mathfrak{h}_{c}$ by eq.(\ref{hprime}), it is evidently
a real form of $\mathfrak{h}^\mathbb{C}$ and  
$\mathfrak{g}=\mathfrak{h}\cap \mathfrak{h}_{c}$. 
Given a decomposition (\ref{Cartan}) with properties (\ref{CartanProp}), $\theta $ as defined in (\ref{thetaact}) is an involutive automorphism.
\end{proof}
The real forms of simple complex Lie algebras are classified in text books \cite{fuchsschwei}. 
The real forms of semisimple or reductive Lie algebras (i.e. semisimple except for abelian factors) can be deduced from them. 

Our main interest in the present paper will be in the following 

\begin{example} $\mathfrak{h}^{\mathbb{C}}= \mathfrak{so}(5,\mathbb{C})$. 

{\em We state the results for the pairs of groups $H\supset G$, $G\neq H$  modulo passage to covering groups.  }

a) $G$ compact: $SO(1,4)\supset SO(4)$ and $SO(2,3)\supset S(O(2)\times O(3))$.

b) $G$ noncompact: $SO(1,4)\supset SO(1,3)$ and $SO(2,3)\supset SO(1,3)$.
\end{example} 
We see that the involutive automorphism is unique for both choices $SO(1,4)$ and $SO(2,3)$ of $H$ if it is demanded that $G$ is noncompact, and $G=SO(1,3)$ in both cases.

If $G$ is the maximal compact subgroup of $H$ then 
$\mathfrak{h}_{c}$ is the compact real form of $\mathfrak{h}^{\mathbb{C}}$ and $\theta$ is known as the Cartan involution of $\mathfrak{h}$. 
The Cartan involution lifts to an involution of $H$ and the split (\ref{Cartan}) leads to a global decomposition, i.e. the elements $h$ of $H$ can be written as 
\begin{equation}\label{Cartangroup}
h=g\exp(X),\quad g\in G,\, X\in\mathfrak{p}.
\end{equation}
The Cartan decomposition generalizes the polar decomposition of matrices.

In ref \cite{Lehmann:jn} it is  shown   that the parametrization (\ref{Cartangroup}) of parallel transporters in extra directions yields Higgs fields $X$ which take their values in the tangent space $\mathfrak{p}$ to the coset space $H/G$. This coset space $H/G$ is a symmetric space in case $G$ is the maximal compact subgroup of $H$. $\mathfrak{p}$ is isomorphic
to the noncompact part of the Lie algebra of $H$.

Here we are interested to accomodate Vierbein fields in an analogous way. Then one has to admit  maximal noncompact subgroups $G$ of $H$ as gauge groups 
in order to get the Lorentz group as unitary gauge group.

Even in this case
one can show that a decomposition on the group level corresponding to eq. (\ref{Cartan}) exists, at least for elements nearby the identity of $H$. We comment 
on global decompositions in the outlook. 
% The proof of existence and uniqueness of such a generalized polar decomposition is due to Lawson[?]. Here we give an alternative proof, see also \cite{Falkmaster}. 
\begin{theo}[Polar decomposition].\label{polardeco} Let $\mc{T}\in\mc{N}\subset H$, where $\mc{N}$ is a sufficiently small neighborhood of the identity in $H$. Suppose that there is an involutive antiautomorphism $g\mapsto g^{\ast}$ of $H$, which passes to an involutive antiautomorphism of the Lie algebra of $H$.

Then $\mc{T}$ can be uniquely represented in the form
\begin{equation}
\mc{T}=\mc{U}\mc{P},
\end{equation}
where $\mc{U}$ and $\mc{P}$ satisfy
\begin{eqnarray}
\mc{U}^{\ast} & = &\mc{U}^{-1}\\
\mc{P}^{\ast} & = & \mc{P} , 
\end{eqnarray}
i.e. the first factor $\mc{U}$ is unitary, the second factor $\mc{P}$ selfadjoint, and both are close to the identity.
\end{theo}
\begin{proof}[
{\sc Proof of Theorem \ref{polardeco}}]\cite{Falkmaster}.
Let $\mc{N}$ be a sufficiently small neighborhood of the identity in $H$ such that for $\mc{T}\in\mc{N}$
we have a unique representation $ \mc{T}^{\ast}\mc{T}=e^{X},X\in\frak{h}$. Then also $ (\mc{T}^{\ast}\mc{T})^{1/2}=e^{X/2}$ and within $\mc{N}$ there are uniquely determined elements
\begin{eqnarray}
\mc{P} & := & (\mc{T}^{\ast}\mc{T})^{1/2}\\
\mc{U} & := & \mc{T}(\mc{T}^{\ast}\mc{T})^{-1/2}
\end{eqnarray}
with $\mc{T}=\mc{U}\mc{P}$.
Clearly, one finds $\mc{P}^{\ast}=\mc{P}$ and $\mc{U}^{\ast}=\mc{U}^{-1}$.
\end{proof}
%%%%%%%%%%%%%%%%%%%%%%%%%%%%%%%%%%%%%%%%%%%%%%%%%%%%%%%%%%%%%
\subsection{$\ast$-operation on parallel transporters}
An involutive automorphism $\theta$ of the holonomy group $H$ suffices to specify a $\ast$-operation on the PTs.
More precisely, we have
\begin{theo}\label{uniquestar}
Given parallel transporters $\mc{T}(C)$ in a vector bundle $\mc{V}$ with fibres $V_{x}\cong V=V_{\hat x}$, suppose that the associated holonomy group $H_{\hat x}\cong H$ is equipped with an involutive automorphism $\theta$.
Then the induced $\ast$-operation $\mc{T}(C)^{\ast}=\theta(\mc{T}(C))^{-1}$ on  PTs along closed paths from $\hat x$ to $\hat x$ extends to a $\ast$-operation on arbitrary PTs. The $\ast$-operation is determined by the conjugacy class of $\theta$, up to active gauge transformations in $H$.  
\end{theo}
\begin{proof}[
{\sc Proof of Theorem \ref{uniquestar}.}] First we prove the existence. 
Given $V_{x},V_{y}$ we let $\mathbb{E}(V_{x},V_{y}):=\lbrace \mc{T}(C)\vert C:x\mapsto y\rbrace$ denote the set  of all parallel transporters along paths $C$ from $x$ to $y$. 
Elements of $\mathbb{E}(V_{x},V_{y})$ can be identified with elements of the holonomy group $H_{\hat x}$ in the following way.
Given $x$ choose a path $C_{x}$ from $x$ to $\hat x$. Let $i_{yx}$ be a map  $i_{yx}:\mathbb{E}(V_{x},V_{y})\rightarrow H_{\hat x}$ defined by
$
i_{yx}(\mc{T}(C)):=\mc{T}(C_{y})\mc{T}(C)\mc{T}(C_{x})^{-1}.
$
Note that $i_{yx}$ is a functor from the category of all parallel transporters on $\mc{M}$ to the holonomy group $H_{\hat x}$, i.e. $i_{zy}i_{yx}=i_{zx}$.

For $\mc{T}(C)\in\mathbb{E}(V_{x},V_{y})$ we define $\mc{T}(C)^{\ast}\in\mathbb{E}(V_{y},V_{x})$ by
\begin{equation}\label{defstar}
\mc{T}(C)^{\ast}:=i_{xy}^{-1}(\theta(i_{yx}(\mc{T}(C))^{-1}).
\end{equation}
Due to the composition law of PTs and the properties of $\theta$ and $i_{yx}$, respectively, the $\ast$-operation satisfies
$(\mc{T}(C_{2})\mc{T}(C_{1}))^{\ast}=\mc{T}(C_{1})^{\ast}\mc{T}( C_{2})^{\ast}$ and $\mc{T}(C)^{\ast\ast}=\mc{T}(C)$.

Next we show that the definition (\ref{defstar}) does not depend on the choice of path $C_{x}$ up to an active gauge transformation. 
Let $C'_{x}$ be another path from $x$ to $\hat x$ and $i_{yx}'$ be the associated map.
Then we get after a short calculation
\begin{equation}
\mc{T}(C)^{\ast'}=S(x)\overline{\mc{T}(C)}^{\ast}S(y)^{-1}
\end{equation}
where $S(z)=\mc{T}(C'_{z})^{-1}\mc{T}(C_{z})\in H_{z}\cong H$ and $\overline{\mc{T}(C)}=S(y)^{-1}\mc{T}(C)S(x)$
\end{proof}
Given the PTs, the definition of the associated vector potential depends on a choice of moving frame. 
 Introducing a moving frame which furnishes bases in $V_{x}$, PTs  get converted to matrices via  
$\mc{T}(C)e_{a}(x)=e_{b}(y)\bs{T}^{b}\,_{a}(C).
$
Assuming differentiability of the connection,  
the vector potential $\bs{B}_{\mu}(x)$ is defined by considering infintesimal paths $b:x\mapsto x+\delta x$
\begin{equation}\label{vecpot}
\bs{T}(b)=\bs{1}-\bs{B}_{\mu}(x)\delta x^{\mu}.
\end{equation}
Since $\bs{T}(C)\in H, \bs{B}_{\mu}(x)$ is in its Lie algebra $\mathfrak{h}$. On the Lie algebra level one has $X^{\ast}=-\theta(X)$. Therefore
\begin{equation}\label{spliti}
\bs{B}_{\mu}(x)=\bs{A}_{\mu}(x)+\bs{E}_{\mu}(x)
\end{equation}
with
\begin{equation}\label{splitii}
\bs{A}_{\mu}(x)=-\bs{A}_{\mu}(x)^{\ast}\in\mathfrak{g}\quad\text{and}\quad\bs{E}_{\mu}(x)=\bs{E}_{\mu}(x)^{\ast}\in\mathfrak{p}.
\end{equation}
Therefore the $\ast$-operation induces a split of the vector potential. We shall later use this to identify a spinorial form of the vierbein and the spin connection as parts of a single de Sitter vector potential. Referring to Example 1
we see that $\bs{A}_{\mu}(x)\in\mathfrak{so}(1,3)$ if $H\cong SO(1,4)$ or $SO(2,3)$ and $G$ is noncompact. We shall examine the transformation properties
of the pieces under gauge transformations in Theorem \ref{trafo} below. The $\ast$-operation acts on infinitesimal PT as follows
\begin{equation}
\boldsymbol{T}(b)^{\ast}:=\boldsymbol{1}-(\bs{E}_{\mu}(x)-\bs{A}_{\mu}(x))\delta x^{\mu}.
\end{equation}
Consequently we find
\begin{equation}\label{nonu}
\bs{T}(b)^{\ast}\bs{T}(b)=\bs{1}-2\bs{E}_{\mu}(x)\delta x^{\mu}\neq\bs{1},
\end{equation}
i.e. $\bs{T}(C)^{\ast}\neq\bs{T}(C)^{-1}$ for nonvanishing $\bs{E}_{\mu}(x)$.
%%%%%%%%%%%%%%%%%%%%%%%%%%%%%%%%%%%%%%%%%%%%%%%%%%%
\begin{theo}\label{theorem4A}
Given the $\ast$-operation on parallel transporters $\mc{T}(C)$, these admit a unique decomposition
\begin{equation}
\mc{T}(C)=\mc{U}(C)\mc{P}(C)\quad \text{for} \quad C:x\rightarrow y
\end{equation}
into a unitary factor $\mc{U}(C)=\mc{U}(C)^{\ast-1}:V_{x}\rightarrow V_{y}$
which obeys the composition law 
\begin{equation}
\mc{U}(C_{2}\circ C_{1})=\mc{U}(C_{2})\mc{U}(C_{1}),\quad\mc{U}(\emptyset)=\text{id}
\end{equation}
and $H_{x}\ni\mc{P}(C):V_{x}\rightarrow V_{x}$ such that $\mc{P}(b)=\mc{P}(b)^{\ast}$ for infinitesimal paths $b$.
Let 
\begin{equation}
\mathfrak{h}_{x}=\mathfrak{g}_{x}+\mathfrak{p}_{x}
\end{equation}
be the split of the Lie algebra $\mathfrak{h}_{x}$ of $H_{x}$ into $\ast$-odd and $\ast$-even parts. This defines $\mathfrak{p}_{x}$ as a real subspace of $End(V_{x})$
and the unitary factor can serve to define parallel transport of elements of $\mathfrak{p}_{x}$ via
\begin{equation}
\label{unitarytransport}
\mathfrak{p}_{x}\ni X\rightarrow \mc{U}(C)X\mc{U}(C)^{-1}\in\mathfrak{p}_{y}
\end{equation}
for arbitrary paths $C$ from $x$ to $y$.
\end{theo}
\begin{proof}[
{\sc Proof of Theorem \ref{theorem4A}.}]
Consider the corresponding parallel transport matrices $\bs{U}(C)$ and $\bs{P}(C)$. For infinitesimal paths $b$ it follows from $\mc{P}(b)=\mc{P}(b)^{\ast}$ and eqs. 
(\ref{vecpot})-(\ref{splitii}) that
 \begin{eqnarray}
\bs{P}(b) & = & \bs{1}-\bs{E}_{\mu}(x)\delta x^{\mu}\label{infiniself}\\ 
\bs{U}(b) & = & \bs{1}-\bs{A}_{\mu}(x)\delta x^{\mu},\label{conventionalptmatrix}
\end{eqnarray} respectively.
Existence and uniqueness of the decomposition  for finite paths $C$ follows from the composition laws by composing $C$ from infinitesimal paths. The explicit formulae
for $\bs{U}(C)$ and $\bs{P}(C)$ are given in Theorem \ref{globalptmatrix} below. 

Finally we prove eq. (\ref{unitarytransport}). It suffices to consider infinitesimal paths $C$ from $x$ to $x+\delta x$. Passing to matrix description we have 
$\bs{U}(C)=\bs{1}-\bs{A}_{\mu}(x)\delta x^{\mu}$ with $\bs{A}_{\mu}(x)\in\mathfrak{g}_{x}$. If $X\in\mathfrak{p}_{x}$ then
\begin{equation}
 \mc{U}(C)X\mc{U}(C)^{-1}=X+[X,\bs{A}_{\mu}(x)]\delta x^{\mu}%\in\mathfrak{p}_{x}
\end{equation}
%because $[\mathfrak{p}_{x},\mathfrak{g}_{x}]\in\mathfrak{p}_{x}$.
and this is $\ast$-even.
\end{proof}
\begin{theo}\label{globalptmatrix}
Given the path $C$ parametrized by $\tau\in[\tau_{f},\tau_{i}]$, write $\bs{U}[\tau_{2},\tau_{1}]$ for the unitary parallel transporters along the piece of $C$ from $C(\tau_{1})$ to $C(\tau_{2})$. Define the covariant line integral
\begin{equation}
\int_{C}\bs{E}_{\mu}(x)Dx^{\mu}:=\int^{\tau_{f}}_{\tau_{i}}\bs{U}[\tau,\tau_{i}]^{-1}\bs{E}_{\mu}(x(\tau))\bs{U}[\tau,\tau_{i}] d\tau.
\end{equation}
Then 
 \begin{equation}
\bs{U}(C)=T\exp\left(-\int_{C}\bs{A}_{\mu}(x)dx^{\mu}\right),
\end{equation}
where $T$ is ordering with respect to the parameter $\tau$and
\begin{equation}\label{finitep}
\bs{P}(C)=T\exp\left(-\int_{C}\bs{E}_{\mu}(x)Dx^{\mu}\right).
\end{equation}
\end{theo}
\begin{proof}[
{\sc Proof of Theorem \ref{globalptmatrix}.}]
The path $C$ can be decomposed into infinitesimal pieces $C=b_{N}\circ\ldots b_{1}, N\rightarrow\infty$.
Inserting the polar decomposition for the infinitesimal pieces, one obtains the formula
\begin{equation}
\bs{T}(C)=\bs{U}(b_{N})\bs{P}(b_{N})\ldots\bs{P}(b_{2})\bs{U}(b_{1})\bs{P}(b_{1}),
\end{equation}
in the limit $N\rightarrow \infty$. The $\bs{P}$-factors can be pushed to the right, using $\bs{P}\bs{U}=\bs{U}\bs{P}'$, where
$\bs{P}':=\bs{U}^{-1}\bs{P}\bs{U}$.
As a result one arrives at formula (\ref{finitep}).
\end{proof}
Active gauge transformations in $H$ induce linear transformations  of moving frames, described by matrices $\bs{S}(x)$, according to
 \begin{equation}
 e_{\alpha}(x)\rightarrow e_{\alpha}'(x)=g_{x}e_{\alpha}(x)=e_{\beta}(x)\bs{S}(x)^{\beta}\,_{\alpha},\quad g_{x}\in H_{x}.
 \end{equation} 
 Under conditions laid down in section 2.1 $\bs{S}(x)\in H$. General parallel transport matrices along paths $C$ from $x$ to $y$ transform according to 
\begin{equation}\label{transformpt}
\bs{T}(C)\mapsto \bs{S}(y)^{-1}\bs{T}(C)\bs{S}(x).
\end{equation} 
 We restrict attention to gauge transformations which are unitary in the sense that $g_{x}^{\ast}=g_{x}^{-1}$.
We call them {\em unitary gauge transformations} for short. The corresponding matrices $\bs{S}(x)$ form subgroups $\bs{G}_{x}$ isomorphic to $G$. For suitably restricted moving frames, the matrix group $\bs{G}_{x}$ is
independent of $x$. By abuse of notation, we write $\bs{S}(x)\in G$ for unitary gauge transformations.
The transformation behaviour of the pieces of the vector potential is given by the next theorem.
\begin{theo}[Tranformation laws]\label{trafo}
Under a unitary gauge transformation $\bs{S}(x)\in G$ the pieces of the vector potential $\bs{B}_{\mu}(x)$ transform according to
\begin{eqnarray}
\bs{E}_{\mu}(x)\rightarrow\bs{E}'_{\mu}(x)&=&\bs{S}^{-1}(x)\bs{E}_{\mu}(x)\bs{S}(x)\label{trafoE},\\
\bs{A}_{\mu}(x)\rightarrow\bs{A}'_{\mu}(x)&=&\bs{S}^{-1}(x)\bs{A}_{\mu}(x)\bs{S}(x)+\bs{S}(x)^{-1}\partial_{\mu}\bs{S}(x).
\end{eqnarray}
\end{theo}
\begin{proof}[
{\sc Proof of Theorem \ref{trafo}.}]
Combining formula (\ref{transformpt}) and (\ref{infiniself}), (\ref{conventionalptmatrix}), we arrive at (\ref{trafoE}).
\end{proof}
We note that $\bs{E}_{\mu}(x)$, which will later be identified with the spinorial form of the vierbein, transforms homogeneously, as it must be.
%%%%%%%%%%%%%%%%%%%%%%%%%%%%%%%%%%%%%%%%%%%%%%%%%%%%%%%%%%%%%%%%%%%%%%%%%%%%%%%%%%%%%%%%%%%%%%%%%%%%
\subsection{$\ast$-representation of the holonomy group}\label{starrep}
 Starting from a principal fibre bundle one has to choose a representation $(\tau, V)$ of the holonomy group $H$ to obtain an associated vector bundle. Conventionally, the representation spaces $V$ are equipped
with a scalar product $\langle\quad,\quad\rangle$ such that the adjoint map with respect to $\langle\quad,\quad\rangle$ is the inverse, i.e.
\begin{equation}\label{unitarystarpresentation}
\langle v,\tau(g)w\rangle=\langle \tau(g)^{-1}v,w\rangle.
\end{equation}
Consequently, one deals with unitary (or pseudo-unitary) representations of the holonomy group.

 In the framework of generalized parallel transporters it is natural to admit also nonunitary representations of $H$ which do not satisfy (\ref{unitarystarpresentation}). More precisely we are interested in situations where the algebraic $\ast$-operation is at the same time the adjoint map between vector spaces with a bilinear or sesquilinear form, respectively. This motivates the following definition 
\begin{defi}
Let $H$ be a group which is equipped with an involutive automorphism $\theta$ and let $g^{\ast}:=\theta(g)^{-1}$. A $\ast$-representation of $H$ is a representation of $H$ given by operators $\tau(g):V\rightarrow V$ where $V$ is a finite-dimensional  real or complex vector space
which is equipped with a nondegenerate bilinear or sesquilinear  form $\langle\,,\,\rangle$ such that
\begin{equation}
\langle v,\tau(g)w\rangle=\langle\tau(g^{\ast})v,w\rangle.
\end{equation}
 \end{defi}
 Note that unitary representations are the special case associated with positive definite forms $\langle\quad,\quad\rangle$ and the trivial automorphism $\theta(g)=g$.
 
 \begin{theo}\label{mastertheorem}Suppose that
the manifold  $\mathcal{M}$ is connected. Given parallel transporters $\mathcal{T}(C)$ 
in a vector bundle $\mathcal{V}$ over $\mathcal{M}$ with fibres $V_{x}\cong V=V_{\hat x}$,
suppose that the induced representation of the holonomy group $H$ on $V$ can be made into a $\ast$-representation by a choice of a bilinear or sesquilinear form, denoted $\langle , \rangle_{\hat x}$, on $V$. 
 
 Then  the fibres  $V_{x}$ can be equipped with  bilinear or sesquilinear forms $\langle,\rangle_{x}$, respectively, such that
 \begin{equation}\label{masterstar}
\langle v,\mathcal{T}(C)w\rangle_{y}=\langle\mathcal{T}(C)^{\ast}v,w\rangle_{x}
\end{equation}
for all paths $C:x\rightarrow y$ between arbitrary sites $x,y\in\mathcal{M}$ and all $v\in V_{y},w\in V_{x}$. 
\end{theo}
 %%%%%%%%%%%%%%%%%%%%%%%%%%%%%%%%%%%%%%%%%%%%%%%%%%%%%%%%%%%%%%%%%%%%%%%
\begin{corol} \label{corol1}
Let $\mc{U}(C)$ be the unitary factor in the decomposition of $\mc{T}(C)$ according to Theorem \ref{theorem4A}.
Under the same assumptions as in Theorem \ref{mastertheorem} we have 
\begin{equation}\label{uc}
\langle\mathcal{U}(C)v,\mathcal{U}(C)w\rangle_{y}=\langle v,w\rangle_{x}
\end{equation}
 for all $x,y\in\mc{M},v,w\in V_{x} $ and all paths $C:x\rightarrow y$.
\end{corol}
\begin{proof}[
{\sc Proof of Theorem \ref{mastertheorem}.}]%Note that eq. (\ref{uc}) is a special case of (\ref{masterstar}).
Given $x$, choose a path $C:x\rightarrow \hat x$. Define the bilinear or sesquilinear form $\langle\quad,\quad\rangle_{x}$ in $V_{x}$ by
\begin{equation}\label{scalardefinition}
\langle v,w\rangle_{x}:=\langle\mathcal{U}(C)v,\mathcal{U}(C)w\rangle_{\hat x},
\end{equation}
%for all $v,w\in V_{x}$.
where $\mc{U}(C)$ are unitary parallel transporters introduced in section
 \ref{polarsection}. 
First  we show that the scalar product does not depend on the choice of the path $C$. To see this let $C'$ be another path
from $x$ to $\hat x$. Then $L=C\circ (-C'):\hat x\rightarrow\hat x$ is a closed path, therefore $\mathcal{U}(L)$ is an element of
the unitary gauge group with $\mathcal{U}(L)^{\ast}=\mathcal{U}(L)^{-1}$. As $\mathcal{U}(-C)\mathcal{U}(C)=\text{id}$ it follows that $\mathcal{U}(C)=\mathcal{U}(C)\mathcal{U}(-C')\mathcal{U}(C')=\mathcal{U}(L)\mathcal{U}(C')$. Consequently,
\begin{eqnarray}
\langle\mathcal{U}(C)v,\mathcal{U}(C)w\rangle_{\hat x}&=& \langle\mathcal{U}(L)\mathcal{U}(C')v,\mathcal{U}(L)\mathcal{U}(C')w\rangle_{\hat x}\nonumber\\
{}&=& \langle\mathcal{U}(C')v,\mathcal{U}(C')w\rangle_{\hat x}.
\end{eqnarray} due to the $\ast$-property of the representation of the holonomy group and the unitarity of the parallel transporters. This proves independence of the choice of $C$.
To prove eq. (\ref{masterstar}) we use the identification of
 parallel transporters $\mc{T}(C)$ for arbitrary paths $C$ with elements of the holonomy group $H$ via the map $i_{yx}$ and the definition (\ref{scalardefinition}). Then the statement (\ref{masterstar}) follows from the assumption that we deal with a $\ast$-representation of $H$ as a short calculation shows
\begin{equation}
\begin{split}
\langle v,\mc{T}(C) w\rangle_{y} & = \langle\mc{U}(C_{y})v,\mc{U}(C_{y})\mc{T}(C)w\rangle_{\hat x}
=\langle\mc{U}(C_{y})v,i_{yx}(\mc{T}(C))\mc{U}(C_{x})w\rangle_{\hat x}\nonumber\\
&=\langle\mc{U}(C_{x})i_{xy}^{-1}(\theta(i_{yx}(\mc{T}(C)))^{-1})v,\mc{U}(C_{x})w\rangle_{\hat x}
= \langle \mc{T}(C)^{\ast},w\rangle_{x}.
\end{split}\end{equation}
This completes the proof of Theorem \ref{mastertheorem}.
 
The corollary follows from eq. (\ref{masterstar}) and unitarity $\mc{U}(C)^{\ast}=\mc{U}(C)^{-1}$ of the parallel transporters associated with the unitary gauge group $G$. 
\end{proof}
In general relativity we are mainly interested in real vector spaces with an indefinite bilinear form. We shall use Theorem \ref{mastertheorem}
with the following
\begin{lemma}\label{kleinlammer}
Suppose the real vector space $V$ carries a representation of the holonomy group $H$
with involutive automorphism $g\mapsto \theta(g)=g^{\ast -1}$. Let $V'$ be the dual space.
Then $V\otimes V'$ can be equipped with a bilinear form to make  it into a $\ast$-representation space.
\end{lemma}
\begin{proof}[
{\sc Proof of Lemma \ref{kleinlammer}.}] Define the bilinear form on $V\otimes V'$ by
\begin{equation} 
\langle v\otimes\xi,w\otimes\chi\rangle:=\xi(w)\chi(v)\qquad\quad v,w\in V,\,\,\xi,\chi\in V'.
\end{equation}
The representation $\tau^{\otimes}(g):V\otimes V'\rightarrow V\otimes V'$ is given by
 \begin{equation}
(\tau^{\otimes}(g))(v\otimes\xi):=\tau(g)v\otimes\tau'(g)\xi,
\end{equation}
where $\tau'$ denotes the representation carried by the dual space $V'$, which is defined as
%The representation of the dual space is given by
\begin{equation}
(\tau'(g)\xi)(v):=\xi(\tau(g^{\ast})v),\qquad v\in V,\xi\in V'.
\end{equation}
A simple calculation yields
\begin{eqnarray}
\langle v\otimes\xi,(\tau^{\otimes}(g))(w\otimes\chi)\rangle &=& \langle(\tau^{\otimes}(g^{\ast})(v\otimes\xi),w\otimes\chi\rangle.
\end{eqnarray}
\end{proof}
%%%%%%%%%%%%%%%%%%%%%%%%%%%%%%%%%%%%%%%%%%%%%%%%%%%%%%%%%%%%%%%%%%%%%%%%%%%
 %%%%%%%%%%%%%%%%%%%%%%%%%%%%%%%%%%%%%%%%%%%%%%%%%%%%%%%%%%%%%%%%%%%%%%%%%%% 
 %%%%%%%%%%%%%%%%%%%%%%%%%%%%%%%%%%%%%%%%%%%%%%%%%%%%%%%%%%%%%%%%%%%%%%%%%%%%%%%%%%%%%%%%%%%%%%%%%%%%%%%%%%%%%%%%%%%%%%%%%%%%%%%%%%%%%%%%%%%%%%%%%%%%%%%%%%%%%%%%%%%%%%%%%%%%%%%%%%%%%%%
\section{Nonunitary parallel transport \\ of Dirac spinors}\label{diracspinors}
Let us now apply our general formalism developed so far to parallel transport of Dirac spinors.
 In the traditional approach to general relativity, one considers the parallel transport $\mc{U}(C):T_{x}\mc{M}\rightarrow T_{y}\mc{M}$
of tangent vectors. It satisfies the unitarity condition (\ref{unitary}), and there is a scalar product $\langle\quad,\quad\rangle_{x}$ in $T_{x}\mc{M}$,
which is determined  by the metric tensor $g_{\mu\nu}(x)$ and which is invariant under parallel transport
\begin{equation}\label{starproperty}
\langle\mc{U}(C)^{\ast}v,w\rangle_{x}=\langle v,\mc{U}(C)w\rangle_{y},\quad \text{for all}\quad v\in T_{y}\mc{M},w\in T_{x}\mc{M}.
\end{equation}

When an pseudo-orthonormal moving frame $(e_{\alpha}(x))$ is chosen, the corresponding vector potential $\bs{A}_{\mu}(x)=(A^{\alpha}\,_{\beta\mu}(x))$
is in the Lie algebra $\mathfrak{so}(1,3)$ of the Lorentz group.
Indices $\alpha,\beta$ are raised and lowered  with the metric tensor
$\eta_{\alpha\beta}=diag(+1,-1,-1,-1)$ in Minkowski space.
Upon introducing Dirac $\gamma$-matrices which obey standard anticommutation relations $\lbrace\bs{\gamma}_{\alpha},\bs{\gamma}_{\beta}\rbrace=2\eta_{\alpha\beta}$, one defines
\begin{equation}
\bs{A}_{\mu}(x):=\frac{1}{8}A^{\alpha\beta}\,_{\mu}(x)[\bs{\gamma}_{\alpha},\bs{\gamma}_{\beta}].
\end{equation}
It can be used to define parallel transporters $\mc{U}(C)$ for Dirac spinors $\Psi(x)\in V_{x}\cong\mathbb{C}^{4}$.
There is a scalar product $\langle \Psi,\Phi\rangle_{x}=\Psi^{\dagger}(x)\bs{\beta}\Phi(x)$ with $\bs{\beta}=\bs{\gamma}^{0}$ in popular representations of the Dirac matrices, which is invariant under the parallel transport,
so that eq.(\ref{starproperty}) holds again. Generically, the  gauge group is the two fold cover $Spin(1,3)$ of the Lorentz group.

Conversely, if the parallel transport of Dirac spinors is given, complex 4-vectors $v$ which can be made from Dirac spinors can also be parallel transported.
It turns out that there exists a subspace of real 4-vector fields which is preserved by parallel transport.
To obtain the parallel transport of tangent vectors $\partial_{\mu}$ to $\mc{M}$, they need to be identified with real
4-vectors $v=(v^{\alpha})$. This requires the vierbein $e^{\alpha}_{\mu}(x)$. Its square gives the metric, $g_{\mu\nu}(x)=e^{\alpha}_{\mu}(x)
e^{\beta}_{\nu}(x)\eta_{\alpha\beta}$.
In this way, general relativity appears as a gauge theory with gauge group $G=SL(2,\mathbb{C})$ 
\cite{Utiyama:1956sy} and an additional field $e^{\alpha}_{\mu}(x)$.

We propose to incorporate the vierbein into a vector potential associated with a gPT
\begin{eqnarray}
 \bs{B}_{\mu}(x)&:=&\frac{1}{2l}e^{\alpha}_{\mu}(x)\bs{\gamma}_{\alpha}
+\frac{1}{8}A^{\alpha\beta}\,_{\mu}(x)[\bs{\gamma}_{\alpha},\bs{\gamma}_{\beta}]\label{vecpo} \\
{} &=:& \bs{E}_{\mu}(x)+\bs{A}_{\mu}(x),
\end{eqnarray}

$l$ has the dimension of a length and will be chosen conveniently later on.
 The corresponding holonomy group is isomorphic to
the two fold cover $Spin(1,4)$ of a de Sitter group.

 The two pieces of the vector potential may be distinguished by their transformation property under a suitable involutive automorphism $\theta$ of the de Sitter algebra, viz
 \begin{equation}
 -\theta(\bs{X})=\bs{\beta}\bs{X}^{\dagger}\bs{\beta}^{-1}
 \end{equation}
where $\bs{X}^{\dagger}$ is the matrix adjoint of $\bs{X}$. 
Defining the $\ast$-operation on the Lie algebra by $X^\ast = -\theta(X)$,
one verifies that
\begin{equation}
\bs{E}_{\mu}(x)^\ast =+\bs{E}_{\mu}(x),\quad \bs{A}_{\mu}(x)^\ast=-\bs{A}_{\mu}(x)
\end{equation}
We note that the automorphism $\theta$ can be implemented within the representation in the sense that that there exists a matrix $\bs{\Theta}$ such that
\begin{equation}
\theta(\bs{X})=\bs{\Theta}\bs{X}\bs{\Theta}^{-1}, \quad viz \quad \bs{\Theta}=-i\bs{\gamma}_{5}.
\end{equation}
The automorphism $\theta$ of the Lie algebra passes to an involutive automorphism of the group and may serve to define a $\ast$-operation on parallel transporters acting on Dirac spinors in the manner described in section 2. 

Let us mention that there is an alternative to the de Sitter group $SO(1,4)$. Actually, it is also possible to choose
\begin{equation}\label{adschoice}
\bs{E}_{\mu}(x):=\frac{1}{2l}e^{\alpha}_{\mu}(x)\bs{\gamma}_{\alpha}\bs{\gamma}_{5}.
\end{equation}
This choice leads to the (anti)-de Sitter group with Lie algebra $\mathfrak{so}(2,3)$. The Majorana condition on Dirac spinors is  invariant under anti- de Sitter parallel transport, while it is not
under de Sitter parallel transport.
Later it will turn out that Weyl fermions can only be accomodated with a holonomy group $H=Spin(2,3)$. Let us remind ourselves that the Majorana condition on Dirac spinors is defined as
\begin{equation}
\psi^{\bs{C}}:=\bs{C}\bar\psi^{t}=\bs{C}\beta^{t}\psi^{c.c.}\overset{!}{=}\psi,
\end{equation}
where $\bs{C}$ denotes the charge conjugation matrix and $c.c$ means complex conjugation. Invariance under parallel transport requires
\begin{equation}\label{invmaj}
(\mc{T}(C)\psi)^{\bs{C}}=\mc{T}(C)\psi^{\bs{C}}.
\end{equation}%%%%%%%%%%%%%%%%%%%%%%%%%%%%%%%%%%%%%%%%%%%%%%%%%%%%%%%%%%%%%%%%%%%%%%%%%%%%%%%%%%%%%%%%%%%%%%%%%%%%%%%%%%%%%%%%%%%%%%%%%%%%%%%%%%%%%%%%%%%%%%%%%%%%%%%%%%%%%%%%%%%%%%%%%%%%%%%%%%%%%%%%%%%%%%%%%%%%%%%%%%%%%%%%%%%%%%%%%%%%%%%%%%%%%%%%%%%%%%%%%%%%%%%%%%%%%%%%%%%%%%%%%%%%%%%%%%%%%%%%%%%%%%%%%%%%%%%%%%%%%%%%%%%%%%%%%%%%%%%%%%%%%%%%%%%%%%%%%
\subsection{ Weyl spinors}\label{weylspinors}
Up to now we assumed that matter was described by Dirac spinors. However, it is possible to define nonunitary parallel transport also for Weyl spinors, assuming $H=Spin(2,3)$.
 These parallel transporters will be real linear but not complex linear. In the following we shall consider lefthanded
spinors $\xi\in V^{(\frac{1}{2},0)}$ for definiteness sake.

Let $\mc{C}$ be the operator of complex conjugation, and define the following real linear transformations of $V^{(\frac{1}{2},0)}$
\begin{eqnarray}
\bs{\rho}_{\alpha}&:=&\frac{1}{l}\bs{\sigma}_{\alpha}\bs{\epsilon}\mathcal{C}\label{generateI}\\
\bs{\rho}_{\alpha\beta}&:=&\bs{\sigma}_{\alpha}\tilde{
\bs{\sigma}}_{\beta}-\bs{\sigma}_{\beta}\tilde{\bs{\sigma}}_{\alpha},\label{generateII}
\end{eqnarray}where $\bs{\epsilon}$ is the antisymmetric tensor in two dimensions and $\bs{\sigma}_{i}$ are Pauli matrices, $\tilde{\bs{\sigma}}_{i}=-\bs{\sigma}_{i}$ and $\tilde{\bs{\sigma}}_{0}=\bs{\sigma}_{0}=\bs{1}$. They are Lorentz covariant in the sense that
\begin{equation}
\bs{S}\bs{\rho}_{\alpha}\bs{S}^{-1}=\bs{\rho}_{\beta}\Lambda^{\beta}\,_{\alpha}(\bs{S})
\end{equation}for $\bs{S}\in SL(2,\mathbb{C})$,
and satisfy the same commutation relations as $\bs{\gamma}_{\alpha}-i\bs{\gamma}_{5}$ and $[\bs{\gamma}_{\alpha},\bs{\gamma}_{\beta}]$,
in particular
\begin{equation}
\lbrack\bs{\rho}_{\alpha},\bs{\rho}_{\beta}\rbrack=-l^{-2}\bs{\rho}_{\alpha\beta}.
\end{equation}
Therefore they generate the Lie algebra $\mathfrak{so}(2,3)$. The vector potential associated with the nonunitary parallel transport of Weyl spinors is
\begin{eqnarray}
\bs{B}_{\mu}(x)&:=&\frac{1}{2}e^{\alpha}_{\mu}\bs{\rho}_{\alpha}+\frac{1}{8}A^{\alpha\beta}\,_{\mu}(x)\bs{\rho}_{\alpha\beta}\\
&=:&\bs{E}_{\mu}(x)+\bs{A}_{\mu}(x).
\end{eqnarray}

In place of eq.(\ref{generateI}), (\ref{generateII}) one could define
\begin{eqnarray}
\bs{\rho}_{\alpha}&:=&\frac{1}{l}i\bs{\sigma}_{\alpha}\bs{\epsilon}\mathcal{C}\\
\bs{\rho}_{\alpha\beta}&:=&\bs{\sigma}_{\alpha}\tilde{\bs{\sigma}}_{\beta}-\bs{\sigma}_{\beta}\tilde{
\bs{\sigma}}_{\alpha}.
\end{eqnarray}
They satisfy the same commutation relations. 
Apparently, it is not possible to accomodate the Lie algebra $\mathfrak{so}(1,4)$ here. The $\ast$-operation is obtained  the involutive anti-automorphism $\theta$
\begin{equation}
\bs{X}^{\ast}=-\theta(\bs{X}):=\epsilon^{-1}\bs{X}^{t}\epsilon
\end{equation}
for $\bs{X}\in\mathfrak{so}(2,3)$. It can be used to identify the two pieces of the vector potential
$\theta(\bs{E}_{\mu}(x))=-\bs{E}_{\mu}(x),\quad\theta(\bs{A}_{\mu}(x))=+\bs{A}_{\mu}(x)$. The automorphism passes to an involutive automorphism of the two fold cover of the anti-de Sitter group.
The elements of the unitary gauge group $G=SL(2,\mathbb{C})$ are characterized by 
\begin{equation}
\theta(g)=g^{-1}\Leftrightarrow g\in SL(2,\mathbb{C}).
\end{equation}
The polar decomposition of parallel transporters along infinitesimal paths $b$ is $\bs{T}(b)=\bs{U}(b)\bs{P}(b)$ with
$\bs{P}(b)=\bs{1}-\bs{E}_{\mu}(x)\delta x^{\mu}$ and $\bs{U}(b)=\bs{1}-\bs{A}_{\mu}(x)\delta x^{\mu}$, respectively.
So far, everything looks very similiar to the Dirac case. But beware: Only the unitary parallel transporters are $\mathbb{C}$-linear. This fact requires some care in  calculations.  For details of the "Weyl-formalism", see the diploma thesis of F. Neugebohrn \cite{Falkmaster}.
%%%%%%%%%%%%%%%%%%%%%%%%%%%%%%%%%%%%%%%%%%%%%%%%%%%%%%%%%%%%%%%%%%%%%%%%%%%%%%%%%%%%%%%%%%%%%%%%%%%%%%%%%%%%%%%%%%%%%%%%%%%%%%%%%%%%%%%%%%%%%%%%%%%%%%%%%%%%%%%%%%%%%%%%%%%%%%%%%%%%%%%%%%%%%%%%%%%%%%%%%%%%%%%%%%%%%%%%%%%%%%%%%%%%%%%%%%%%%%%%%%%%%%%%%%%%%%%%%%%%%%%%%%%%%%%%%%%%%%%%%%%%%%%%%%%%%%%%%%%
\section{Generalized metricity}\label{genmetr}
We now wish to define parallel transport of tangent vectors to $\mc{M}$, and a metric on $\mc{M}$ such that the length of tangent vectors is invariant under
parallel transport.
The idea is simple. The metric comes from the vierbein part of the de Sitter vector potential, and the parallel transport comes from the unitary factor
in the de Sitter parallel transporters. We must explain what means `` comes from'', and show that the announced properties hold.

With a view towards generalizations beyond general relativity, we beginn the discussion without assuming that the holonomy group is de Sitter.
Let us assume that the parallel transport $\mc{T}(C)$ on some space of spinors $V_{x}\ni\psi(x)$ is defined. We write $V_{x}'$ for the space of linear maps
\begin{equation}
\alpha:V_{x}\rightarrow\mathbb{C},\quad v\rightarrow \alpha(v).
\end{equation}
Parallel transport of fibers $V_{x}$ passes to parallel transport of fibers $V_{x}'$ in a canonical way,
\begin{equation}
(\mc{T}(C)\alpha)(v):=\alpha(\mc{T}(C)^{\ast}v)
\end{equation}
for $C:x\rightarrow y$, and therefore also to $V_{x}\otimes V_{x}'$.
The same is true of the unitary parallel transport defined in Theorem \ref{theorem4A}. The space $End(V_{x})$ of linear maps $V_{x}\rightarrow V_{x}$ is canonically isomorphic to $V_{x}\otimes V_{x}'$ because $v\otimes f$ defines a map $u\mapsto vf(u)$.
Referring to the last part of Theorem \ref{theorem4A}, we see that the vierbein,
which is $\ast$-even, maps elements $t=t^\mu \partial_\mu$ of tangent space 
$T_x \mathcal{M}$ into $\mathfrak{p}_x \subset End(V_x)$, 
\begin{equation}
t^\mu \bs {E}_{\mu}(x)\in\mathfrak{p}_{x}\subset V_{x}\otimes V_{x}'
\end{equation} 
and its unitary parallel transport is defined
\begin{equation}
 t^\mu \bs{E}_{\mu}(x)\mapsto t^\mu \bs{U}(C)\bs{E}_{\mu}(x)\bs{U}(C)^{-1}\in\mathfrak{p}_{y}.
\end{equation}
Combining Theorem \ref{mastertheorem} and Lemma \ref{kleinlammer}, $V_{x}\otimes V_{x}'$ gets equipped with a nondegenerate bilinear form such that the $\ast$-property (\ref{masterstar}) holds.
Consider the unitary factors $\mc{U}(C)$ in the generalized polar decompositionof PT's. %Unitarity reads$\mc{U}(C)^{\ast}=\mc{U}(C)^{-1}$, consequently
Corollary \ref{corol1} asserts that
\begin{equation}
\langle\mc{U}(C)w,\mc{U}(C)z\rangle_{y}=\langle w, z\rangle_{x}, \label{metricPres}
\end{equation}
where $\langle\quad,\quad\rangle_{x}$ denotes the bilinear form on $V_{x}\otimes V_{x}'$.
Since $t^\mu \bs{E}_{\mu}(x)\in\mathfrak{p}_{x}\subset V_{x}\otimes V_{x}'$, the 1-form $\bs{E}(x):=\bs{E}_{\mu}(x)dx^{\mu}$ defines a map
\begin{equation}
\bs{E}(x) :T_{x}\mathcal{M}\rightarrow \mathfrak{p}_{x}\subset V_{x}\otimes V'_{x}
\end{equation}
from the tangent space of $\mc{M}$ to a real subspace $\mathfrak{p}_{x}$ of the span of vectors $v\in V_{x}\otimes V_{x}'$.
This defines a bilinear form on $T_{x}\mc{M}$, i.e. a metric $\langle\partial_{\mu},\partial_{\nu}\rangle_{x}=g_{\mu\nu}(x)$ via
\begin{equation}\label{metric}
g_{\mu\nu}(x)=\langle\bs{E}(\partial_{\mu}),\bs{E}(\partial_{\nu})\rangle_{x}.
\end{equation}
If the range of the map obeys
\begin{equation}\label{survierbein}
\bs{E}(x)\left\lbrack T_{x}\mathcal{M}\right\rbrack=  \mathfrak{p}_{x}
\end{equation}
then $\bs{E}(x)$ identifies $\mathfrak{p}_{x}$ with the tangent space $T_{x}\mc{M}$ and unitary parallel transport in the space of vectors $v\in\mathfrak{p}_{x}$
passes to a  parallel transport of tangent vectors which is metric preserving
by eq.(\ref{metricPres}). 

It is of interest to study also the degenerate case
\begin{equation}
E(x)\left\lbrack T_{x}\mathcal{M}\right\rbrack=:W_{x}\subset V_{x}\otimes V'_{x},\quad   W_{x}\neq\mathfrak{p}_{x}.
\end{equation}
We shall return to this case in section 7.
 %%%%%%%%%%%%%%%%%%%%%%%%%%%%%%%%%%%%%%%%%%%%%%%%%%%%%%%%%%%%%%%%
\subsection{Metricity in general relativity}
In general relativity, the condition (\ref{survierbein}) is satisfied. Indeed, the subspace $\mathfrak{p}_{x}$ of the de Sitter Lie algebras $\mathfrak{so}(1,4)$
or $\mathfrak{so}(2,3)$ is a 4-dimensional real vector space (spanned by Dirac matrices $\bs{\gamma}_{\alpha}$ or $\bs{\gamma}_{\alpha}\bs{\gamma}_{5}$, respectively), and the metric is nondegenerate  % only 
if the image of $T_{x}\mc{M}$ under $\bs{E}(x)$ also has real dimension 4. 

The metric defined by eq. (\ref{metric})  does not have the customary dimension, though. Therefore we replace it by
\begin{equation}
g_{\mu\nu}(x)=l^{2}\langle\bs{E}(\partial_{\mu}),\bs{E}(\partial_{\nu})\rangle_{x},
\end{equation}
where $l$ is the standard of length which was introduced in section 3 and which will be conveniently chosen later on. In other words, the
true vierbein is $\bs{e}_{\mu}(x)=l\bs{E}_{\mu}(x)$.

In the case of $H=Spin(2,3)$ and using the Weyl spinor formalism of section 3.1., the scalar product $\langle\quad,\quad\rangle_{x}$ in $
V_{x}\otimes V_{x}'\cong End(V_{x})$ is given by a trace and we have 
\begin{equation}
\bs{E}(x)=\frac{1}{2l}e^{\alpha}_{\mu}(x)\bs{\sigma}_{\alpha}\bs{\epsilon}\mc{C}dx^{\mu}.
\end{equation}
Therefore we arrive   at
\begin{eqnarray}
tr(\bs{E}(\partial_{\mu})\bs{E}(\partial_{\nu}))&=&\frac{1}{4l^{2}}tr(e^{\alpha}_{\mu}(x)e^{\beta}_{\nu}(x)\bs{\sigma}_{\alpha}\tilde{\bs{\sigma}}_{\beta})\\
&=& \frac{1}{2l^{2}}e^{\alpha}_{\mu}(x)e^{\beta}_{\nu}(x)\eta_{\alpha\beta}\\&=& \frac{1}{2l^{2}}g_{\mu\nu}(x).
\end{eqnarray}
 As expected, we get the customary metric tensor of general relativity.
 
 In the case of Dirac spinors, the discussion proceeds in the same way (using $\text{tr}(\bs{\gamma}_{\alpha}\bs{\gamma}_{\beta})=4\eta_{\alpha\beta}$).

%%%%%%%%%%%%%%%%%%%%%%%%%%%%%%%%%%%%%%%%%%%%%%%%%%%%%%%%%%%%%%%%%%%%%%%%%%%%%%%%%%%%%%%%%%%%%%%%%%%%%%%%%%%%%%%%%%%%%%%%%%%%%%%%%%%%%%%%%%%%%%%%%%%%%%%%%%%%%%%%%%%%%%%%%%%%%%%%%%%%%%%%%%%%%%%%%%%%%%%%%%%%%%%%%%%%%%%%%%%%%%%%%%%%%%%%%%%%%%%%%%%%%%%%%%
\section{Gravity Actions}\label{gravityaction}
Let us now turn to the formulation of an action for gravity within the general framework developed so far. We shall consider in the following section a generalized gauge theory with  holonomy group $H=Spin(1,4)$.
We introduce the matrix-valued de Sitter field strength 2-form which is associated with the de Sitter vector potential $\bs{B}$
\begin{equation}
\bs{F}=d\bs{B}+\bs{B}\wedge\bs{B}.
\end{equation}

Using the split (\ref{vecpo}) of the generalized vector potential we find
\begin{equation}
\bs{F}= 
  \bs{F}^{\mc{U}}+\bs{E}\wedge\bs{E}+\bs{T}\label{gravity}
\end{equation}
Here 
 $\bs{F}^{\mc{U}}=d\bs{A}+\bs{A}\wedge\bs{A}$ is the Lorentz curvature and $\bs{T}=\du \bs{E}=d
\bs{E}+\bs{E}\wedge\bs{A}+\bs{A}\wedge\bs{E}$ is the torsion, $\du$ being the 
exterior covariant  derivative associated with the spin connection.

Under the $\ast$-operation the field strength $\bs{F}$ decomposes in two different parts.
The odd part of the field strength is
$
\bs{F}^{-}:=\frac{1}{2}\left(\bs{F}-\bs{F}^{\ast}\right)=\bs{F}^{\mc{U}}+\bs{E}\wedge\bs{E},
$
whereas the even part is the torsion tensor
$
\bs{T}=\frac{1}{2}\left(\bs{F}+\bs{F}^{\ast}\right).
$
We propose the following action for gravity
\begin{equation}\label{uniqueaction}
S=\frac{1}{g^{2}}\int\text{tr}(\bs{F}\wedge\bs{F}^{\ast}\bs{\Theta})=-\frac{1}{g^{2}}\int\text{tr}(\bs{F}\wedge\bs{F}\bs{\Theta})
\end{equation}
where $\bs{F}$ is the de Sitter field strength in the Dirac spinor representation and $\bs{\Theta}=-i\bs{\gamma}_{5}$ is the implementation of the outer automorphism $X \mapsto -X^\ast$ of the de Sitter Lie algebras  in the Dirac representation so that
\begin{equation}
\bs{F}^{\ast}\bs{\Theta}=-\bs{\Theta}\bs{F}.
\end{equation}
We chose to demand $\bs{\Theta}^2=-1$, where $-1$ is the nontivial element of the center of $Spin(1,4)$. As a result, $\bs{\Theta}$ defines a complex 
structure of the Lie algebra. 
 T. Grimm \cite{grimm} pointed out a possible connection with  recent work of Hitchin \cite{Hitchin:2004ut,Hitchin:2000jd} where complex structure also plays a crucial role. 

The equality of both expressions for $S$ follows from the cyclicity of the trace.

$\frac{1}{g^{2}}$ is a dimensionless constant which will be identified with $\frac{3}{16\pi G\Lambda}$ in a moment, where $G$ is the Newton  constant and $\Lambda$ the cosmological constant.

Note that the action involves a kind of supertrace,
\begin{equation}
\text{Tr}\bs{\omega}:=\text{tr}(\bs{\omega}\bs{\Theta}) \ .
\end{equation}
Before deriving the field equations, let us establish that the action (\ref{uniqueaction}) actually yields the Einstein-Palatini action with cosmological constant plus a topological term. 
Inserting the split (\ref{gravity}) of the de Sitter field strength 
the terms involving $\bs{T}$, which is a linear combination of $\gamma$-matrices, vanish because of properties of traces of products of $\gamma$-matrices. 
We get
\begin{equation}\label{einsteinpalatini}
S=\frac{i}{g^{2}}\int tr(\bs{F}^{\mc{U}}\wedge\bs{F}^{\mc{U}}\bs{\gamma}_{5})+\frac{i}{2g^{2}l^{2}}\int tr(\bs{e}\wedge\bs{e}\wedge\bs{F}^{\mc{U}}\bs{\gamma}_{5})+\frac{i}{16g^{2}l^{4}}\int tr(\bs{e}\wedge\bs{e}\wedge\bs{e}\wedge\bs{e}\bs{\gamma}_{5})
\end{equation}
The first term is topological because $\bs{F}^{\mc{U}}$ maps the two irreducible 2-dimensional representation spaces for the Lorentz group in themselves,
 and $\bs{\gamma}_{5}$  restricts to $\pm 1$ on these spaces. The other two terms in (\ref{einsteinpalatini}) are a spinorial rewriting of the Einstein-Palatini  action with cosmological constant $\Lambda$ (see the Appendix)
\begin{equation}\label{standardaction}
S_{E.P.}=\frac{1}{16\pi G}\int d^{4}x\det e (R+2\Lambda)
\end{equation}
 if we identify
\begin{equation}\label{constants}
G:=\frac{l^{2}g^{2}}{16\pi},\quad\Lambda:=\frac{3}{l^{2}},\quad \frac{1}{g^{2}}=\frac{3}{16\pi G\Lambda}.
\end{equation}
Remarkably, we get a realistic value for the energy density $\rho_{\Lambda}=\frac{\Lambda}{8\pi G}$  if we identify the length $l$ with the ultimate infrared cutoff, namely the Hubble constant.
In this case $g^{2}\sim 10^{-120}$ is a tiny constant.

In the de Sitter vector potential, the part $\bs{E}=\frac{1}{2l}e^{\alpha}\bs{\gamma}_{\alpha}$ and $\bs{A}$ are independent.
Variation of the Einstein Palatini action with respect to $\bs{E}$ and $\bs{A}$ yields the Einstein field equations, here with a cosmological constant $\Lambda$, $\bs{E}\wedge\bs{F}^{\mc{U}}+\bs{F}^{\mc{U}}\wedge\bs{E}+2\bs{E}\wedge\bs{E}\wedge\bs{E}=0$ and vanishing of the torsion. A more careful derivation of the field equations will be given in a moment and confirms the result.

Let us comment on the uniqueness properties of this action, given that we do not have a proper Hodge $\star$-operator. In the context of generalized gauge theory we have actually two parallel transporters
$\mc{T}(C)$ and $\mc{T}(C)^{\ast-1}$ along a path $C$. Correspondingly, there exist two covariant derivatives, which we denote by $\dt$ and $\dts$ respectively and which are conjugate under 
$\bs{\Theta}$ in the sense of eq.(\ref{thetaConjug}), and two field strengths.
But the two field strengths are simply $\bs{F}$ and $-\bs{F}^{\ast}$. Therefore the ambiguity of which field strength to take is irrelevant in view of the equality $(\ref{uniqueaction})$. The only other candidate for an 
$\mc{F}^{\mc{T}}$ squared action which is not purely topological  would be 
\mbox{$\frac{1}{g^{2}}\int\text{tr}(\bs{F}\wedge\bs{F}^{\ast})$.}
 This would not lead to the Einstein field equations, but instead to the condition of vanishing torsion, as we shall see later on.

Regarding $S$ as a function of the de Sitter vector potential $\bs{B}$ its variation with respect to $\bs{B}$ yields
\ba
\delta_{\bs{B}}S&=&
-\int \text{tr}(\delta\bs{F}\wedge \bs{F}\bs{\Theta}
+ \bs{F} \wedge \delta\bs{F}\bs{\Theta}) \\
&=& -\int \text{tr} \left( \dt \delta \bs{B} \wedge (\bs{F}\bs \Theta
+ \bs{\Theta} \bs{F})\right)
\ea
Let us pretend for a moment that $\delta \bs{B} $ is an arbitrary element of the Clifford algebra generated by the Dirac matrices. Actually it is not so, but we show below that it makes no difference.
Upon partial integration, the vanishing of $\delta_{\bs{B}}S$ 
 yields the field equations 
\begin{equation}\label{einstein} 
\dt (\bs{F}\bs{\Theta} -
\bs{F}^{\ast}\bs{\Theta}
)=0,
\end{equation}
or, equivalently, $\dts (\bs{F} - \bs{F}^{\ast })=0$.

An obvious solution of (\ref{einstein}) is
\begin{equation}\label{solution}
\bs{F}=0
\end{equation}
(\ref{solution}) does not yield a trivial solution, but instead one gets
\begin{eqnarray}
\bs{F}^{\mc{U}}&=&-\bs{E}\wedge\bs{E}\label{desittersolution}\\
\bs{T}&=&0
\end{eqnarray} 
Note that the solution of (\ref{desittersolution}) describes a de Sitter universe with cosmological constant $\Lambda$.

Next we split $\dt (\bs{F}\bs{\Theta} - \bs{F}^{\ast}\bs{\Theta} )$ into linearly independent pieces which must vanish separately.  Inserting the above expression for $\bs{F}^{-}$ the field equation reads
\begin{equation}
0= \du (\bs{E}\wedge\bs{E})+\bs{E}\wedge\bs{F}^{\mc{U}}+\bs{F}^{\mc{U}}\wedge\bs{E}+2\bs{E}\wedge\bs{E}\wedge\bs{E} \ . \label{einstein2}
\end{equation}
 We note that   $\bs{E}\wedge\bs{F}^{\mc{U}}+\bs{F}^{\mc{U}}\wedge\bs{E}+2\bs{E}\wedge\bs{E}\wedge\bs{E}$ involves products of odd numbers of $\gamma$-matrices, and $\du (\bs{E}\wedge\bs{E})$ involves even numbers. So they are linearly independent. As a result we obtain the following two independent 
equations
\begin{eqnarray}
\bs{E}\wedge\bs{F}^{\mc{U}}+\bs{F}^{\mc{U}}\wedge\bs{E}+2\bs{E}\wedge\bs{E}\wedge\bs{E}&= & 0\label{einsteineq}\\
\du (\bs{E}\wedge\bs{E}) &= & 0 \ . \label{torsionfree}
\end{eqnarray}
(\ref{einsteineq}) is just the spinorial form of the Einstein field equation
with a cosmological constant. (\ref{torsionfree}) is 
well known from the analysis of the Palatini variational principle as 
the condition for vanishing torsion.

Notice that in the action (\ref{uniqueaction}) there is no reason to assume invertibility of the vierbein.
Also the field equations make perfect sense without assuming a invertible vierbein.

Let us finally dispense with the presumption that $\delta \bs{B} $ is an 
arbitrary element of the Clifford algebra. In actual fact, it must be an element of the Lie algebra $so(1,4)$, hence of the form 
$$ \delta \bs{B} = \frac 1 2 \delta E^\alpha \bs{\gamma}_\alpha 
+ \frac 1 8 \delta A^{\alpha \beta} [\bs{\gamma}_\alpha, \bs{\gamma}_\beta] \
$$
The only nonvanishing trace of a product of Dirac-matrices multiplied with 
$\bs{\Theta} = -i \bs{\gamma}_5$ is 
$-i \text{tr} \bs{\gamma}_\alpha \bs{\gamma}_\beta \bs{\gamma}_\gamma 
\bs{\gamma}_\delta \bs{\gamma}_5 = 4 \epsilon_{\alpha \beta \gamma \delta }$. From the stationarity of $S$ under $\delta \bs{B}$ of the above form, we may therefore only conclude that the contributions to the right hand side (r.h.s.) of 
eq.(\ref{einstein2}) must vanish which are proportional to totally antisymmetric products of two or three Dirac matrices. But a short calculation reveals
 that all the terms on the r.h.s. of eq.(\ref{einstein2}) are of this form. Therefore eq.(\ref{einstein2}) must hold as it stands.

There have been speculations about the existence of  a low energy phase in which gravity is described by general relativity and a high energy phase which is governed 
by a purely topological theory with vanishing vierbein.
 We  have argued in refs \cite{Lehmann:2003jn,Lehmann:2003jh} that the emergence of nonunitary PTs may result from a RG-flow. 
It is tempting to ask whether there is a connection between the emergence of nonunitary PTs and the transition from a topological phase
to a low energy phase governed by general relativity.
 
%Furthermore, let us mention that recently Freidel and Starodubtsev used a topological de Sitter theory to argue that general relativity might be pertubatively renormalizable \cite{Freidel:2005ak}.

Turning to the conventional massless Dirac matter action, one can show that it
 also may be rewritten in a form which does not need an inverse vierbein. Writing
$D^{\mc{U}}:=D^{\mc{U}}_{\mu}dx^{\mu}=(\partial_{\mu}+\bs{A}_{\mu})dx^{\mu}$ and $\bs{E}(x):=\frac{1}{2l}e^{\alpha}\,_{\mu}(x)\bs{\gamma}_{\alpha}dx^{\mu}$ we have
\begin{equation}\label{polymatter}
S_{M}=\int\,d^{4}x\det(e^{\alpha}_{\mu})\bar\psi e^{\mu}_{\alpha}\bs{\gamma}_{\alpha}D^{\mc{U}}_{\mu}\psi\propto l^{3}\int \bar\psi\bs{e}\wedge\bs{e}\wedge\bs{e} \bs{\Theta}\wedge D^{\mc{U}}\psi.
\end{equation}
%where we have used
It is natural in the spirit of this paper to  replace $D^{\mc{U}}$ by $D^{\mc{T}}=d+\bs{B}$ in (\ref{polymatter}). This adds a tiny mass term
\begin{equation}\label{polymatter2}
\begin{split}S_{M}=\int\,d^{4}x\det(e^{\alpha}_{\mu})\bar\psi e^{\mu}_{\alpha}\bs{\gamma}_{\alpha}D^{\mc{T}}_{\mu}\psi\propto l^{3}\int \bar\psi\bs{e}\wedge\bs{e}\wedge\bs{e} \bs{\Theta}\wedge D^{\mc{T}}\psi&=\nonumber\\
l^{3}\int \bar\psi\bs{e}\wedge\bs{e}\wedge\bs{e} \bs{\Theta}\wedge D^{\mc{U}}\psi+l^{4}\int \bar\psi\bs{e}\wedge\bs{e}\wedge\bs{e}\wedge\bs{e} \bs{\Theta}\psi.
\end{split}
\end{equation}
Finally, let us consider the action
\begin{equation}\label{alternateaction}
S=\int \text{tr}(\bs{F}\wedge\bs{F}^{\ast}).
\end{equation}
Variation with respect to $\bs{B}$ results in
\begin{equation}
\dt\bs{F}^{\ast}=0.
\end{equation}
Because of the Bianchi identity $\dt \bs{F}=0$, this is equivalent 
to $ 0=\frac 12 \dt (\bs{F}+\bs{F}^\ast)=\dt \bs{T} = \du \bs{T} + \bs{E}\wedge \bs{T} -
  \bs{T}\wedge \bs{E}$. This splits as before into 
\begin{eqnarray}
\bs{E}\wedge \bs{T} -
  \bs{T}\wedge \bs{E} &=&0\\
\du \bs{T}&=&0.
\end{eqnarray}
The first equation is the same as $\du (\bs{E}\wedge\bs{E})=0$.
We conclude that the action (\ref{alternateaction}) leads to the condition of vanishing torsion, but not to the Einstein field equations. A term proportional
 to it could be added to the action which we proposed. Classically, all these actions yield the same field equations. But quantum fluctuations would differ.
 %%%%%%%%%%%%%%%%%%%%%%%%%%%%%%%%%%%%%%%%%%%%%%%%%%%%%%%%%%%%%%%%%%%%%%%%%%%%%%%%%%%%%%%%%%%%%%%%
%%%%%%%%%%%%%%%%%%%%%%%%%%%%%%%%%%%%%%%%%%%%%%%%%%%%%%%%%%%%%%%%%%%%%%%%%%%%%%%%%%%%%%%%%%%%%%%
%%%%%%%%%%%%%%%%%%%%%%%%%%%%%%%%%%%%%%%%%%%%%%%%%%%%%%%%%%%%%%%%%%%%%%%%%%%%%%%%%%%%%%%%%%%%%%%%%%%%%%%%%%%%%%%%%%%%%%%%%%%%%%%%%%%%%%%%%%%%%%%%%%%%%%%%%%%%%%%%%%%%%%%%%%%%%%%%%%%%%%%%%%%%%%%%%%%%%%%%%%%%%%%%%%%%%%%%%%%%%%%%%%%%%%%%%%%%%%%%%%%%%%%%%%%%%%%%%%%%%%%%%%%%%%%%%%%%%%%%%%%%%%%%%%%%%%%%%%%%%%%%%%%%%%%%%%%%%%%%%%%%%%%%%%%%%%%%%%%%%%%%%%%%%%%%%%%%%%%%%%%%%%%%%%%%%%%%%%%%%%%%%%%%%%%%%%%%%%%%%%%%%%%%%%%%%%%%% 
\section{Classical equations of motion}\label{classical}
Up to now our discussion was based on the parallel transport of Dirac spinors. It turns out, somewhat surprisingly, that also classical massive particles can be treated within the de Sitter framework. Let us recall that in general relativity a classical point particle is described by its four-vector $u^{\alpha}(\tau):=\frac{dx^{\alpha}(\tau)}{d\tau}$, where $(u^{\alpha})$ are the components with respect to an orthonormal basis.  Denoting the Lorentz covariant derivative by $D^{\mc{U}}$, the equation of motion is
\begin{equation}\label{motion}
\left(\frac{D^{\mc{U}}}{d\tau}u(\tau)\right)^{\alpha}:=\frac{d}{d\tau}u^{\alpha}+A^{\alpha}\,_{\beta\mu}u^{\beta}u^{\mu}=0.
\end{equation}
Equivalently, the dynamics is determined by
\begin{equation}
(\mc{U}(C)u(\tau))^{\alpha}=u(\tau+\delta\tau)^{\alpha}.
\end{equation}
Within the de Sitter framework the same equation describes the motion, one has just to replace $\mc{U}(C)$ by $\mc{T}(C)$
\begin{equation}
(\mc{T}(C)u(\tau))^{\alpha}=u(\tau+\delta\tau)^{\alpha}.
\end{equation}
To see this, one has  to define the de Sitter parallel transport of four-vectors.
We exploit the fact that the four-velocity is a future-directed, timelike four-vector, i.e. $u^{0}>0$ and $\eta(u,u):=\eta_{\alpha\beta}
u^{\alpha}u^{\beta}>0$.
We identify $u$ with an equivalence class of future-directed, lightlike five-vectors and define their de Sitter parallel transport.

Let $(v^{\alpha})$ be the   components of   a future-directed, timelike four-vector, i.e.
\begin{equation}
(v^{\alpha})\in\mathcal{C}_{+}:=\lbrace (w^{\alpha})\vert w^{0}>0\,\,\,\text{and}\,\,\,\eta(w,w)>0\rbrace.
\end{equation}
Define 
\begin{equation}
v_{4}:=\vert\vert v\vert\vert_{1,3}=(v_{\alpha}\eta^{\alpha\beta}v_{\beta})^{\frac{1}{2}} \quad\text{and}\quad\eta^{44}=-1.
\end{equation}
Now we can define a future-directed, lightlike five-vector $(v_{\alpha},v_{4})$  
\begin{equation}
(v_{\alpha},v_{4})\in\mathcal{C}^{(1,4)}_{+}:=\lbrace (w_{\alpha},w_{4})\vert w_{0}>0,
\vert\vert(v_{\alpha},v_{4})\vert\vert_{1,4}:=v_{\alpha}\eta^{\alpha\beta}v_{\beta}+v_{4}\,\eta^{44}\,v_{4}=0\rbrace.
\end{equation}
Evidently, multiplication with a positive real number yields again an element in $\mathcal{C}^{(1,4)}_{+}$.
The resulting equivalence classes are elements ("rays") of a real projective space
 \begin{equation}
\mathbb{P}^{1,3}:=\lbrace\lbrack v\rbrack\vert v\in\mathcal{C}^{1,4}_{+}\rbrace,
\end{equation}
where
\begin{equation}
\lbrack v\rbrack:=\lbrace w\vert w=\lambda v,\lambda>0\rbrace.
\end{equation}
$u$ satisfies $\vert\vert u\vert\vert_{1,3}=1$. Every four-velocity determines uniquely a ray. Since every ray contains a vector with $u^{4}=1$,
the converse is also true.
Elements of $SO(1,4)$ act as pseudorotations in $\mathbb{R}^{5}$ and map lightlike vectors to lightlike ones.
The parallel transport of rays can be defined as 
\begin{equation}
\mathcal{T}(C)[v]:=\lbrack\mathcal{T}(C)v\rbrack.
\end{equation}
We define a vectorial (as opposed to spinorial) form of the de Sitter vector potential by
\begin{eqnarray}
B^{\alpha \beta}\,_{\mu}(x)&=& A^{\alpha \beta}\,_{\mu}(x) \\
B^{\alpha 4}\,_{\mu}(x)&=& l^{-1}e^\alpha\,_{\mu}(x) = -B^{4 \alpha}\,_{\mu}(x).
\end{eqnarray}
Now the parallel transport of five-vectors may be defined as
\begin{eqnarray}
(\mc{T}(b)v)^{a}&:=& v^{a}(x+\delta x)-[\partial_{\mu}v(x)^{a}+B^{a}\,_{b\mu}v(x)^{b}]\delta x^{\mu}\\
&=
:&v(x+\delta x)^{a}-(D_{\mu}v(x))^{a}\delta x^{\mu}.
\end{eqnarray}
with implied summation over $b=0,...,4$ and 5-dimensional metric $(+,----)$.

%This may be rewritten  
%\begin{eqnarray}
%\left(\mathcal{T}(b)v(\tau)\right)^{\alpha}&=& v(\tau+\delta\tau)^{\alpha}-%\left(\frac{D^{\mc{U}}}{d\tau}v\right)^{\alpha}+\frac{1}{\mathfrak{l}}v^{4}u(\tau)^{\alpha}\delta\tau.\label{rewr%itten}
%\end{eqnarray}

Since $u^{\alpha}$ is a four-vector, we have $u_{\alpha}\eta^{\alpha\beta}u_{\beta}=1$, and $u$ can be identified with the five-vector 
  $(u_{\alpha},1)$. Due to (\ref{motion}) %and (\ref{rewritten})
  it follows that
\begin{equation}
\left(\mathcal{T}(C)u(\tau)\right)^{\alpha}=u(\tau+\delta\tau)^{\alpha}(1+\frac{1}{l}\delta\tau).
\end{equation}
Since $(1+\frac{1}{l}\delta\tau)>0$ we get
\begin{equation}
\mathcal{T}(C)\lbrack u(\tau)\rbrack=\lbrack u(\tau+\delta\tau)\rbrack.
\end{equation}
  As a result, the equation of motion takes the following form in terms of the3 de Sitter covariant derivative
\begin{equation}
\frac{D}{d\tau}\lbrack u(\tau)\rbrack=0. 
\end{equation}
For a representative of the equivalence class, the equation of motion is
\begin{equation}\label{dSttercov}
\left(\frac{D^{\mc{T}}}{d\tau}u(\tau)\right)^{a}=u(\tau)^{a}.
\end{equation}
Obviously, (\ref{dSttercov}) is de Sitter covariant.
Recall that the energy-momentum tensor of a  classical point particle is determined by its four-velocity
\begin{equation}
T_{\alpha\beta}(x)=m\int d\tau\,(-g(x))^{-1/2}\delta^{4}(x-C(\tau))u_{\alpha}(\tau)u_{\beta}(\tau).
\end{equation}
Thus, also the de Sitter parallel transport of $T_{\alpha\beta}$ is defined.%%%%%%%%%%%%%%%%%%%%%%%%%%%%%%%%%%%%%%%%%%%%%%%%%%%%%%%%%%%%%%%%%%%%%%%%%%%%%%%%%%%%%%%%%%%%%%%%%%%%%%%%%%%%%%%%%%%%%%%%%%%%%%%%%%%%%%%%%%%%%%%%%%%%%%%%%%%%%%%%%%%%%%%%%%%%%%%%%%%%%%%%%%%%%%%%%%%%%%%%%%%%%%%%%%%%%%%%%%%%%%%%%%%%%%%%%%%%%%%%%%%%%%%%%%%%%%%%%%%%%%%%%%%%%%%%%%%%%%%%%%%%%%%%%%%%%%%%%%%%%%%%%%%%%%%%%%%%%%%%%%%%%%%%%%%%%%%%%
%%%%%%%%%%%%%%%%%%%%%%%%%%%%%%%%%%%%%%%%%%%%%%%%%%%%%%%%%%%%%%%%%%%%%%%%%%%
%%%%%%%%%%%%%%%%%%%%%%%%%%%%%%%%%%%%%%%%%%%%%%%%%%%%%%%%%%%%%%%%%%%%%%%%%%%
\section{Summary and Outlook}
In this paper we have introduced a framework which is more general than conventional gauge theories.
Following the strategy of lessening what is assumed a priori, we proposed to admit parallel transporters which violate the unitarity condition. In particular
we proposed to abandon the a priori assumption of metricity in general relativity.

We have shown that a $\ast$-operation on PTs can be employed as a substitute, and this $\ast$-operation turned out to be unique, if the holonomy group 
is a de Sitter group. Starting from a single de Sitter gauge field $\bs{B}_{\mu}(x)$, the $\ast$-operation determines a split of $\bs{B}_{\mu}(x)$
into a metric connection and a vierbein which gives the metric.

 An action principle has been formulated and variation with respect to $\bs{B}_{\mu}(x)$ yields both the Einstein field equations and the condition of
vanishing torsion. 

Let us outline some directions for future applications and developments.
In the spirit of this paper it is natural to require that the $\ast$-operation is not fixed a priori but is itself a dynamical variable
that obeys local equations of motion.

Some such proposals were discussed by Smolin and Starodubtsev \cite{Smolin:2003qu}.
 One may also wish to consider group- or involutive-automorphism-valued fields 
$\bs{\Theta} (x)$.
%In such a theory it may be determined dynamically, if one obtains a purely topological theory (correponding to $\bs{\Theta}=\bs{1}$) or gravity.Similarly, the surviving unitary gauge group and thus the signature of the space time metric would ensue dynamically.

In section 4 we pointed out that it may be interesting to study degenerate vierbein fields, i.e.
\begin{equation}
\bs{E}(x)[T_{x}\mc{M}]=W_{x},\quad\mathfrak{p}_{x}=W_{x}\oplus W^{\perp}_{x}.
\end{equation} 
Degenerate vierbein fields may provide a way to unify gravity and the other interactions in a way which is different from the Kaluza-Klein approach with a nondegenerate metric in the higher dimensional space.
More precisely, in this case the unitary vector potential splits into
\begin{equation}
\left(
\begin{array}{cc}
\tilde{\bs{A}}_{\mu}(x) & \tilde{\bs{E}}_{\mu}(x)\\
\tilde{\bs{E}}^{\ast}_{\mu}(x) & \bs{\Phi}_{\mu}(x)
\end{array}\right),
\end{equation}
with $\tilde{\bs{A}}_{\mu}(x)\in Lie\tilde G$ where $\tilde G$ is the subgroup of $G$ which leaves $W_{x}$ invariant. Under $\tilde G$ $
\tilde{\bs{A}}_{\mu}(x)$ transforms like a vector potential, wheras $\tilde{\bs{E}}_{\mu}(x)$ transforms homogeneously and $\bs{\Phi}_{\mu}(x)$ behaves like a scalar.
 $\tilde{\bs{A}}_{\mu}(x)$ can be regarded as the spin connection of gravity which defines parallel transport of vectors in $W_{x}$.

Assume that $\tilde G\otimes K\subset G$, where $K$ is a compact Lie group. 
Then $\bs{\Phi}_{\mu}(x)$ may be interpreted as a gauge field associated with the inner gauge group $K$ and $\tilde{\bs{E}}_{\mu}(x)$
as a sort of Higgs field.
%Note that this approach is very different from the Kaluza Klein paradigm.

Let us emphasize that general relativity on a differentiable manifold does not push Einstein's principle to a logical conclusion. 
While it is basic for general relativity to eliminate the a priori notion of a straight line, the a priori given differential structure of the manifold amounts to specifying 
what is a straight line in the infintesimally small.
 This motivates to study generalized gauge theories on discrete manifolds which are graphs.
Dimakis and M\"uller-Hoissen \cite{Dimakis:1994bq,Dimakis:1994qq,Dimakis:1995dr} have shown that a differential calculus and geometry can be formulated on graphs without further a priori structure wich would substitute for a differentiable structure.
There exists a vector potential but it is no longer in the Lie algebra of a gauge group. As a result, there is no natural possibility to demand 
%that\begin{equation}\mc{T}(-C)=\mc{T}(C)^{-1}.\end{equation}
%A paper on these issues is in preperation.
unitarity of parallel transporters. This provides another motivation for abandonning unitarity of parallel transporters. 

In the discrete context the question of the global existence of a polar 
decomposition generalizing theorem \ref{polardeco} also arises. Such polar decompositions exist for a general class of Lie-semigroups with involutive anti-automorphisms $g \mapsto g^\ast$ \cite{Lawson:1994}, but are not known for groups
if $G$ is not compact. This suggests that also the assumption of invertibility
of PTs is not totally natural on discrete manifolds. This may have physical consequences, see the discussion at the end of section \ref{holonomy}.

Finally we mention that Higgs fields can appear as parts of generalized PTs in extra dimensions in a novel way such that exponential mass hierarchies appear
when the local gauge symmetry is spontaneusly broken by a Higgs mechanism.
More about %the last two issues
discrete theories and Higgs fields
 can be found in \cite{ThorstenDiss}.
%%%%%%%%%%%%%%%%%%%%%%%%%%%%%%%%%%%%%%%%%%%%%%%%%%%%%%%%%%
\subsubsection*{Acknowledgements} It is a pleasure to thank Falk Neugebohrn for
many discussions. We are also indebted to Mathias de Riese for helpful 
discussions on polar decompositions.
T.P. wishes to thank the Deutsche Forschungsgemeinschaft
for financial support through the Graduiertenkolleg {\it Zuk\"unftige Entwicklungen in der Teilchenphysik}. 
%%%%%%%%%%%%%%%%%%%%%%%%%%%%%%%%%%%%%%%%%%%%%%%%%%%%%%%%%%%%%%%%%%%%%%%%%%
\section{Appendix}
Here we would like to show that (\ref{uniqueaction}) turns into the standard action for gravity with a cosmological constant (\ref{standardaction}).

The field strength associated to the spin connection is
\begin{equation}\label{curvature}
\bs{F}^{\mc{U}}_{\mu\nu}=\frac{1}{8}R^{\alpha\beta}\,_{\mu\nu}[\bs{\gamma}_{\alpha},\bs{\gamma}_{\beta}].
\end{equation}
The Riemann-Christoffel curvature tensor corresponding to the spin connection can be constructed in the following way
\begin{equation}
R^{\rho\sigma}\,_{\mu\nu}=e^{\rho}\,_{\alpha}e^{\sigma}\,_{\beta}R^{\alpha\beta}\,_{\mu\nu}.
\end{equation}
The Ricci-tensor is defined by 
\begin{equation}
R^{\sigma}\,_{\nu}=R^{\rho\sigma}\,_{\rho\nu}. 
\end{equation}
With (\ref{curvature}) and $\bs{E}=\frac{1}{2l}e^{\alpha}\,_{\mu}\bs{\gamma}_{\alpha}dx^{\mu}$ the action(\ref{uniqueaction}) results in (neglecting the topological term)
\begin{equation}
-\frac{1}{4g^{2}l^{2}}\int e^{\alpha}\,_{\mu}e^{\beta}\,_{\nu}R^{\gamma\delta}\,_{\rho\sigma}
\epsilon_{\alpha\beta\gamma\delta}\epsilon^{\mu\nu\rho\sigma}d^{4}x-\frac{1}{4g^{2}l^{4}}\int
e^{\alpha}\,_{\mu}e^{\beta}\,_{\nu}e^{\gamma}\,_{\rho}e^{\delta}\,_{\sigma}\epsilon_{\alpha\beta\gamma\delta}\epsilon^{\mu\nu\rho\sigma}d^{4}x,
\end{equation}
where we utilized
\begin{equation}
dx^{\mu}\wedge dx^{\nu}\wedge dx^{\rho}\wedge dx^{\sigma}=\epsilon^{\mu\nu\rho\sigma}d^{4}x
\end{equation}
and
\begin{equation}
\text{tr}(\bs{\gamma}_{\alpha}\bs{\gamma}_{\beta}\bs{\gamma}_{\gamma}\bs{\gamma}_{\delta}\bs{\gamma}_{5})=4i\epsilon_{\alpha\beta\gamma\delta}.
\end{equation}
Employing
\begin{equation}
\det e=-\frac{1}{4!}e^{\alpha}\,_{\mu}e^{\beta}\,_{\nu}e^{\gamma}\,_{\rho}e^{\delta}\,_{\sigma}\epsilon_{\alpha\beta\gamma\delta}\epsilon^{\mu\nu\rho\sigma}
\end{equation}
we finally arrive at
\begin{equation}
S=-\frac{1}{g^{2}}\int\text{tr}(\bs{F}\wedge\bs{F}\bs{\Theta})
=
\frac{1}{g^{2}l^{2}}\int d^{4}x\det e (R+\frac{6}{l^{2}}).
\end{equation}
If we make the identification (\ref{constants}), we obtain (\ref{standardaction})

%\begin{equation}
%\epsilon_{abcd}\epsilon^{\mu\nu\rho\sigma}e^{a}\,_{\mu}e^{b}\,_{\nu}=-2!\det(e)e^{\rho)\,_{[c}e^%{\sigma}\,_{d]}
%\end{equation}

\bibliographystyle{plain}
\bibliography{LiteraturVZ}
%%%%%%%%%%%%%%%%%%%%%%%%%%%%%%%%%%%%%%%%%%%%%%%%%%%%%%%%%%%%%%%%%%%%%%%%%%%%%%%%%%

%%%%%%%%%%%%%%%%%%%%%%%%%%%%%%%%%%%%%%%%%%%%%%%%%%%%%%%%%%%%%%%%%%%%%%
\end{document}